\documentclass[aps,prl,twocolumn,showpacs,amsmath,amssymb,preprintnumbers,10pt,footinbib]{revtex4-1}
\usepackage{epsfig}
\usepackage{amsfonts}
\usepackage{amssymb}
\usepackage{enumitem}
\usepackage{amsthm}
\usepackage{subfig}
\usepackage{bm}
\usepackage{hyperref}
\usepackage{mathrsfs}
\usepackage{bbm}
\usepackage{cancel}
\usepackage[dvipsnames]{xcolor}
\usepackage{accents}
\usepackage{relsize}
\usepackage{float, slashed, graphicx, amssymb, amsmath}
\usepackage{shuffle}
\usepackage{tikz}
\usetikzlibrary{hobby,calc,shapes, positioning, backgrounds,trees, decorations, decorations.markings, decorations.pathmorphing, arrows, shapes.geometric, snakes}
\usepackage{tikz-feynman}
\tikzfeynmanset{compat=1.1.0}
\tikzfeynmanset{
graviton/.style={circle, draw=green!60, fill=green!5, very thick, minimum size=7mm}
}
\tikzfeynmanset{
codot/.style={/tikz/shape=circle,
/tikz/fill=red,/tikz/minimum size=0.1cm,/tikz/inner sep=1.8pt}
}
\tikzfeynmanset{sb/.style=
{/tikz/shape=circle,
/tikz/fill=black,
 /tikz/minimum size=0.3cm,/tikz/inner sep=1.pt
 } }
\tikzfeynmanset{myblob/.style=
{/tikz/shape=ellipse,
/tikz/fill=red,
 /tikz/minimum width=0.5cm,
 } }
 \tikzfeynmanset{myblobFF/.style=
{/tikz/shape=circle,
/tikz/fill=black,
 /tikz/minimum width=0.25cm,
 } }
 \tikzfeynmanset{myblob2/.style=
{/tikz/shape=rectangle,
/tikz/fill=black,
 /tikz/minimum width=0.1cm,/tikz/inner sep=1.8pt
 } }
  \tikzfeynmanset{fp/.style=
{/tikz/shape=rectangle,
/tikz/fill=red,
 /tikz/minimum width=0.1cm,/tikz/inner sep=1.8pt
 } }
  \tikzfeynmanset{myvertex/.style=
{/tikz/shape=circle,
/tikz/fill=black,
 /tikz/minimum width=0.05cm,
 } }
\tikzset{box/.pic={\filldraw[fill=black]  (0,0) circle (2.5pt);
				   \filldraw [fill=black] (0.5,0) circle (2.5pt);
			       \draw [line width=5pt] (0,0) -- (0.5,0);}}

\tikzset{wiggle/.style={decorate, decoration=snake}}

 \tikzfeynmanset{HV/.style=
{/tikz/shape=circle,
/tikz/fill={rgb:black,1;white,2},
 /tikz/minimum size=0.01cm,/tikz/inner sep=1.8pt
 } }

\usepackage[T1]{fontenc} 

\newcommand{\tr}{\text{tr}}

\newcommand{\Id}{\mathbbm{I}}
\newcommand{\cop}{\Delta}
\def\sc#1{\overline{#1}}
\makeatletter
\newcommand \UPlus {\mathop {\operator@font \uplus }\limits }
\makeatother
\makeatletter
\newcommand \Bigcup {\mathop {\operator@font \bigcup }\limits }
\makeatother
  \def\LabelNote#1{}
 \def\Label#1{\label{#1}%
  \smash{\hbox to\phipt{\raise1ex\hbox{\tiny[#1]}\hss}}}
  
  \def\Cdot{{\cdot}}
  
\def\nn{\nonumber}

\newcommand{\widebar}{\overline}

\newcommand{\white}{\color{white}}

\def\spa#1.#2{\left\langle#1\,#2\right\rangle}
\def\spb#1.#2{\left[#1\,#2\right]}

\def\be{\begin{equation}}
\def\ee{\end{equation}}
\def\bea{\begin{eqnarray}}
\def\eea{\end{eqnarray}}  
\allowdisplaybreaks

\newcommand{\graphset}{\mathsf{R}}  
\newcommand{\npre}{\mathcal{N}}

\newcommand{\FF}{q}
\newcommand{\A}{m}
\def\ck#1{\mathsf{K}_{#1}}

\begin{document}

\preprint{
	QMUL-PH-22-22
}

\title{Kinematic Hopf algebra for amplitudes and form factors}

\author{Gang Chen$^1$}
\email{g.chen@qmul.ac.uk}
\author{Guanda Lin$^{1, 2}$}
\email{linguandak@pku.edu.cn}
\author{Congkao Wen$^1$}
\email{c.wen@qmul.ac.uk}
\affiliation{$\mbox{}^{1}$Centre for Theoretical Physics, Department of Physics and Astronomy, 
Queen Mary University of London,\\ Mile End Road, London E1 4NS, United Kingdom}
\affiliation{
CAS Key Laboratory of Theoretical Physics, Institute of Theoretical Physics, Chinese Academy of Sciences, Beijing, 100190, China}

\begin{abstract}

We propose a kinematic algebra for the Bern-Carrasco-Johansson (BCJ) numerators of tree-level amplitudes and form factors in Yang-Mills theory coupled with bi-adjoint scalars. The algebraic generators of the algebra contain two parts: the first part is simply the flavour factor of the bi-adjoint scalars, and the second part that maps to non-trivial kinematic structures of the BCJ numerators obeys extended quasi-shuffle fusion products. The underlying kinematic algebra allows us to present closed forms for the BCJ numerators with any number of gluons and two or more scalars for both on-shell amplitudes and form factors that involve an off-shell operator. The BCJ numerators constructed in this way are manifestly gauge invariant and obey many novel relations that are inherited from the kinematic algebra.  

\end{abstract} 


\maketitle

\section{Introduction}
Gauge and gravity theories play a central role in our understanding of physical phenomena. 
The double copy relation \cite{Bern:2008qj,Bern:2010ue,Bern:2019prr}, which was inspired by Kawai-Lewellen-Tye relations \cite{Kawai:1985xq} in string theory, reveals deep relations between them.
The most critical step of the Bern-Carrasco-Johansson (BCJ) double copy prescription  \cite{Bern:2008qj,Bern:2010ue,Bern:2019prr} is to realise the colour-kinematics duality for gauge theory amplitudes, where the kinematic numerators (also called the BCJ numerators) satisfy the same Jacobi relations as the colour factors.
The colour-kinematics duality discloses the delicate perturbative structures of amplitudes in a large number of gauge theories  \cite{Bargheer:2012gv,Broedel:2012rc,Boels:2013bi,Chiodaroli:2013upa,Bern:2013yya,Johansson:2014zca,Chiodaroli:2014xia,Johansson:2015oia,Chiodaroli:2015rdg,Johansson:2017srf,Chiodaroli:2018dbu}, effective theories \cite{Chen:2013fya,Cheung:2016prv,Carrasco:2016ldy,Mafra:2016mcc,Carrasco:2016ygv,Elvang:2018dco,Carrasco:2019yyn,Carrasco:2021ptp,Chi:2021mio,Menezes:2021dyp,Bonnefoy:2021qgu,Menezes:2021dyp} and can be also extended to form factors 
\cite{Boels:2012ew,Yang:2016ear,Lin:2020dyj, Lin:2021kht,Lin:2021qol,Lin:2021lqo,Li:2022tir}.
It has led to remarkable insights and tremendous progress in the comprehension of amplitudes in both gauge theory and gravity. 

An important approach to study the colour-kinematics duality is to consider the underlying algebraic structures. 
Different versions of kinematic algebras have been realised in a variety of arenas, e.g. the self-dual Yang-Mills (YM) \cite{Monteiro:2011pc}, non-linear sigma model \cite{Cheung:2016prv}, maximally-helicity-violating (MHV) and next-to-MHV sectors of YM theory \cite{Chen:2019ywi,Chen:2021chy}, and Chern-Simons theory \cite{Ben-Shahar:2021zww}.

It was recently found \cite{Brandhuber:2021bsf,Brandhuber:2021kpo} that in the heavy-mass effective theory (HEFT) a quasi-shuffle Hopf algebra \cite{hoffman2000quasi,Blumlein:2003gb,aguiar2010monoidal,hoffman2017quasi,fauvet2017hopf} perfectly depicts the colour-kinematics duality structure in the theory. There are three important ingredients in that algebra: a) the generators as heavy source currents,  b) the fusion product merging two lower-point currents and c) the mapping rule turning the abstract algebraic element to concrete expressions. 
Compact closed expressions of the BCJ numerators for amplitudes of gluons coupled with two heavy particles (as well as pure gluon amplitudes after taking the decoupling limit) were obtained. 
However, there are some restrictions in this prescription: the number of massive particles has to be two,  the heavy-mass limit is required and the physical meaning of the currents and the fusion product is unclear. 
To use the kinematic Hopf algebra to study general amplitudes, we need to circumvent these restrictions. 

In this letter, we present an extended version of the kinematic Hopf algebra which leads to closed-form expressions for amplitudes of any number of scalars without the requirement of the heavy limit. 
More importantly, the physical meaning of the new algebra is transparent: the generators are physical states and the fusion product corresponds to interaction vertices. This understanding results in a universal description of both scattering amplitudes and form factors, where the latter involve certain off-shell gauge-invariant operators. 

In particular, we will consider the YM-scalar theory with a bi-adjoint $\phi^3$ interaction. This particular theory has played a vital role in the study of colour-kinematics duality and double copy \cite{Chiodaroli:2013upa,Chiodaroli:2014xia,Cachazo:2014nsa,Cachazo:2014xea,Chiodaroli:2017ngp} (see further comments in the discussion section). 
The scalars have an identical mass $m$ \footnote{One can simply take $m$ to be zero if a massless scalar theory is required.} and bear colour and flavour indices, denoted as $I$ and $a$ respectively.
We will consider both on-shell amplitudes and form factors with the operator $\operatorname{Tr}(\phi^2)=\sum_{a,I}(\phi^{a,I})^2$ \footnote{The form factors are defined as \unexpanded{$$
    \mathcal{F}(1,2, \ldots, n) = \int d^{D} x\, e^{-i q \cdot x}\langle 1\, 2  \ldots  n |\operatorname{Tr}(\phi^2)| 0\rangle \, ,$$}
where external on-shell states are labelled by $1,2, \ldots, n$, which can be gluons or scalars, and the operator $\operatorname{Tr}(\phi^2)$ carries an off-shell momentum $q=\sum_{i}p_i$. }. 
Unless otherwise stated,   amplitudes/form factors in this paper refer to the colour-ordered amplitudes/form factors and carry flavour indices. Also, we only focus on the single trace ones in the flavour sector of the bi-adjoint scalars, from which one may obtain multi-trace amplitudes using the transmutation operators in \cite{Cheung:2017ems}. 

\section{General framework from  kinematic Hopf algebra}
This section provides a systematic approach constructing amplitudes and form factors via the kinematic Hopf algebra. Unlike the usual Feynman diagram computations, the resulting expressions are manifestly gauge invariant and extremely compact; furthermore, they obey the colour-kinematics duality. The main ingredients for our approach are {\it algebra generators, fusion products and the evaluation map}, which we will describe below. 

The first ingredient is the algebraic generator for  single-particle external states. There are two types of single-particle generators
\begin{equation} \label{eq:genratorK}
  \ck i =
    \begin{cases}
      T^{(i)}_{(i)} 
   \qquad   & \text{for gluons}\\
      T^{(\mathbf\underline i)} t^{a_i} 
     \qquad   &   \text{for scalars}
    \end{cases},  
\end{equation}
where the $T$ represents the kinematic part and $t^{a_i}$ denotes the flavour group generator for the scalars.

Then we combine these single-particle generators together via the fusion product. For now let us consider the simplest examples of (i) the fusion of a single scalar state and a gluon state becoming a two-particle state; and (ii) the subsequent fusion of such a two-particle state fusion with a gluon state into some three-particle states as 
\begin{align}\label{eq:fusionproducteg}
     \ck 1 \star \ck 2&= T^{(1)}t^{a_1}\star T^{(2)}_{(2)}=T^{(12)}_{(2)}t^{a_1}\, , \nn\\
     (\ck 1\star \ck 2) \star \ck 3&=
     T^{(12)}_{(2)}t^{a_1}\star T^{(3)}_{(3)}\\
     &=(-T^{(123)}_{(23)}+T^{(123)}_{(2),(3)}+T^{(123)}_{(3),(2)})t^{a_1}\, , \nn
\end{align} 
where 
on the RHS of these  equations, we have the multi-particle generator
\begin{align}\label{eq:multipart}
	T^{(\alpha)}_{(\tau_1),\cdots, (\tau_r)} t^{a_i}\cdots t^{a_j}\, ,
\end{align}
in which the superscript $\alpha$ denotes the order of the particles in performing the fusion product while the subscript denotes the partition of the gluons, and the product of $t^{a_i}$ from eq.~\eqref{eq:genratorK} composes the flavour structure. 
Notably, the fusion product  is associative: $X\star (Y\star Z)=  (X\star Y)\star Z$ for arbitrary generators $X,Y,Z$. 


The last piece of the construction is the evaluation map $\langle \bullet \rangle$ which is a linear map from an algebra generator to a gauge-invariant expression appearing in actual amplitudes. 
More details on the fusion product and the explicit expressions for the evaluation map will be given later. 

We now show how to use the three ingredients above to obtain tree-level amplitudes and form factors. This can be achieved by giving  the fusion product a physical meaning. The interaction vertices in the Lagrangian usually involve commutator of fields. In our algebraic language, such commutators are exactly the commutators of the fusion products, $[X, Y]=X\star Y {-}Y\star X$. 
More concretely, we express the amplitude as a sum of cubic graphs, and regard each cubic graph as a nested commutator, given the above correspondence between interaction vertices and commutators. 
As an example, for the cubic graph 
\begin{equation}\label{eq:gamma}
\begin{tikzpicture}[baseline={([yshift=-0.0ex,xshift=-0ex]current bounding box.center)}]\tikzstyle{every node}=[font=\small]    \begin{feynman}
    \vertex (a){$\overline{n}$};
     \vertex [above=0.6cm of a](b)[dot]{};
     \vertex [left=0.6cm of b](c);
     \vertex [left=0.22cm of b](c23);
     \vertex [above=0.13cm of c23](v23)[dot]{};
    \vertex [above=.4cm of c](j1){$1$};
    \vertex [right=.7cm of j1](j2){$2\cdots$};
    \vertex [right=0.5cm of j2](j3){$~~n{-}1$};
   	 \diagram*{(a) -- [thick] (b),(b) -- [thick] (j1),(v23) -- [thick] (j2),(b)--[thick](j3)};
    \end{feynman}  
  \end{tikzpicture}\, ,
  \end{equation}
we can interpret the vertices in the graph with commutators of the generators.  The commutators are performed in an ordering $1, 2, \cdots, n{-}1$ and lead to the corresponding nested commutator
\begin{align}
    \widehat\npre([1,2,\cdots, n{-}1])=[\cdots[\ck 1,\ck 2], \cdots, \ck{n{-}1}]\,.
\end{align}
Then the contribution to amplitude is obtained by taking the evaluation map and combining with the propagators  
\begin{align}\label{eq:leftcontribution}
	{\langle\widehat\npre([1,2,\cdots, n{-}1])\rangle\over d_{[1,2,\ldots, n{-}1]}}\, ,
\end{align}
where $d_{[1,2,\ldots, n{-}1]}$ denotes the product of propagators that is associated with the graph. 

Note that for convenience,  we set the $n$-th particle to be a scalar, labelled by $\overline{n}$(characterised by the bar). And in the algebraic construction above, we do not need the generator $\ck n$ since $p_{n}$ can be removed by the momentum conservation. 
For a generic cubic diagram, the contribution takes a similar form as eq.~\eqref{eq:leftcontribution}, except that the commutator structure is determined by the specific cubic graph. As a result, the amplitude is expressed as
\begin{equation} \label{eq:amp}
\begin{aligned}
  \mathcal{A}(\sigma,\overline{n})&=\sum_{\Gamma\in \graphset_\sigma} {\langle \widehat\npre(\Gamma) \rangle \over d_{\Gamma}} \,\,\, \begin{tikzpicture}[baseline={([yshift=-0.8ex]current bounding box.center)}]\tikzstyle{every node}=[font=\small]    
   \begin{feynman}
    \vertex (a)[]{$\overline n$};
     \vertex [above=0.7cm of a](b)[HV]{$\Gamma$};
     \vertex [left=0.6cm of b](c);
    \vertex [above=.4cm of c](j1){$\sigma(1)$};
    \vertex [right=.6cm of j1](j2){$\cdots$};
    \vertex [right=0.6cm of j2](j3){$~~\sigma(n{-}1)$};
   	 \diagram*{(a) -- [thick] (b),(b) -- [thick] (j1),(b)--[thick](j3)};
    \end{feynman}  
  \end{tikzpicture}\\
   &=\sum_{\Gamma\in R_{\sigma}^{(2)}} 
  {\langle[\widehat\npre(\Gamma_a), \widehat\npre(\Gamma_b)]\rangle \over d_{\Gamma_a}d_{\Gamma_b}}  \begin{tikzpicture}[baseline={([yshift=-0.8ex]current bounding box.center)}]\, \tikzstyle{every node}=[font=\small]    
   \begin{feynman}
    \vertex (a)[]{$\sc{n}$};
     \vertex [above=0.6cm of a](b)[dot]{};
     \vertex [left=0.9cm of b](c);
     \vertex [left=0.42cm of b](cL);
     \vertex [above=0.33cm of cL](vL)[HV]{\tiny$\Gamma_1$};
     \vertex [right=0.83cm of vL](vR)[HV]{\tiny$\Gamma_2$};
     \vertex [above=0.6cm of vL](t1){$\sigma_{1}$};
     \vertex [above=0.6cm of vR](t2){$~~\sigma_{2}$};
    \vertex [above=1.1cm of c](j1){};
    \vertex [right=.8cm of j1](j2){};
     \vertex [right=.2cm of j2](j3){};
    \vertex [right=0.8cm of j3](j4){};
   	 \diagram*{(a) -- [thick] (b),(b)--[thick](vL) -- [thick] (j1),(vL) -- [thick] (j2),(vR)--[thick](j3),(b)--[thick](vR)--[thick](j4)};
    \end{feynman}
  \end{tikzpicture}\, ,
\end{aligned}
\end{equation}
where $\graphset_{\sigma}$ represents the cubic diagrams that respect the ordering $\sigma$, the $\graphset_{\sigma}^{(2)}$ denotes all the inequivalent graphs with colour ordering $\sigma$ and two components as cubic graphs $\Gamma_{1},\Gamma_{2}$ and $d_{\Gamma}$ denotes the products of the propagators associated with a given cubic graph $\Gamma$. 
Importantly, by construction the numerators $\langle \widehat\npre(\Gamma) \rangle$ obey Jacobi relations and they are precisely the BCJ numerator of the graph $\Gamma$. 

We will then consider the colour-kinematics duality in form factors \cite{Lin:2021pne}. 
Let us first present the construction then explain the significance shortly.
Graphically, the difference is to replace the interaction vertex involving the $n$-th scalar in eq.~\eqref{eq:amp} with a colourless operator ${\rm Tr}{(\phi^2)}$. 
Here, we assign a fusion product rather than a commutator to the operator. The form factor also has a novel representation that is very analogous to eq.~\eqref{eq:amp},
\begin{align}\label{eq:ff}
 \mathcal{F}_{{\rm Tr}{(\phi^2)}}(\sigma)=&\sum_{\Gamma\in \graphset_{\sigma}^{(2)}}
 {\langle\widehat\npre(\Gamma_{1})\star\widehat\npre(\Gamma_{2})\rangle \over d_{\Gamma_{1}} d_{\Gamma_{2}}} \begin{tikzpicture}[baseline={([yshift=-0.8ex]current bounding box.center)}]\tikzstyle{every node}=[font=\small]    
   \begin{feynman}
    \vertex (a)[]{$q$};
     \vertex [above=0.6cm of a](b)[fp]{};
     \vertex [left=0.8cm of b](c);
     \vertex [left=0.42cm of b](cL);
     \vertex [above=0.33cm of cL](vL)[HV]{\tiny$\Gamma_{1}$};
     \vertex [right=0.8cm of vL](vR)[HV]{\tiny$\Gamma_{2}$};
      \vertex [right=0.62cm of vL](vm){$~$};
     \vertex [above=0.6cm of vL](t1){$\sigma_{1}$};
     \vertex [above=0.6cm of vR](t2){$\sigma_{2}$};
    \vertex [above=1.0cm of c](j1){};
    \vertex [right=.6cm of j1](j2){};
     \vertex [right=.25cm of j2](j3){};
    \vertex [right=0.6cm of j3](j4){};
   	 \diagram*{(a) -- [blue, ultra thick] (b),(b)--[thick](vL) -- [thick] (j1),(vL) -- [thick] (j2),(vR)--[thick](j3),(b)--[thick](vR)--[thick](j4)};
    \end{feynman}  
  \end{tikzpicture} \, .
\end{align}
Finally, $d_{\Gamma_i}$ ($i=1,2$) is the product of propagators in each cubic graph $\Gamma_i$, including the propagator connecting $\Gamma_i$ with the operator (i.e. the red-box vertex).

Importantly, when comparing eq.~\eqref{eq:amp} and \eqref{eq:ff}, we have
\begin{equation*}
    \begin{tabular}{c|c|c}
         & operator vertex & cubic vertex \\\hline
      colour factor   & single trace & structure constant  \\\hline
      algebraic rule &  $X\star Y$ & $[X, Y]$
    \end{tabular}
\end{equation*}
Note that the structure constant is essentially a commutator. 
Then it is understandable to establish the equivalence between the algebraic rule and the physical colour structure. See more evidence in the Supplementary Material, including the extension to form factors of operators like $\tr(\phi^h)$ with $h>2$. 

In the above, we have sketched the algebraic framework and how to obtain the physical observables from it. In the next section, we will spell out the details of the construction. 

\section{Explicit realisation: Fusion product and mapping rule}
We first explain the fusion-product rule at length, which is a non-abelian generalisation of the previous quasi-shuffle product \cite{Brandhuber:2021bsf}. 

As given in eq.~\eqref{eq:multipart}, the generators are in general products of the kinematic part and the flavour part. 
These two parts are commutative and can be treated separately in the fusion product: 
(i) The fusion products of the flavour part are simply the product of the standard Lie algebra generators, which is generally not an abelian product; 
(ii) The kinematic part obeys the non-abelian quasi-shuffle product
\begin{align}\label{eq:stuffleExplicit}
   & T^{(\alpha)}_{(\tau_1), \ldots ,(\tau_r)}\star T^{(\beta)}_{(\omega_1), \ldots ,(\omega_s)} \nn\\
    &= \sum_{ \genfrac{}{}{0pt}{}{\scriptstyle \pi\lvert_{\tau} =\{(\tau_1), \ldots ,(\tau_r)\}}{\scriptstyle \pi\lvert_{\omega} =\{(\omega_1), \ldots ,(\omega_s)\} } }
    (-1)^{t-r-s} T^{(\alpha\beta)}_{(\pi_1), \ldots ,(\pi_t)}\, .
\end{align}
where $\pi|_{\tau}$ (or $\pi|_{\omega}$) means a restriction to the elements of $\pi$ in $\tau$ (or $\omega$), e.g.~$\{(235),(4),(678)\}|_{\{2,3,4,8\}}=\{(23),(4),(8)\}$.
For example, 
\begin{align}
	T^{(12)}_{(1),(2)}\star T^{(345)}_{(34)}=&-T^{(12345)}_{(1),(234)}-T^{(12345)}_{(134),(2)}+T^{(12345)}_{(1),(2),(34)}\nn\\
	&+T^{(12345)}_{(1),(34),(2)}+T^{(12345)}_{(34),(1),(2)}\,.
\end{align}
Compared with the fusion product rules for the amplitudes in HEFT \cite{Brandhuber:2021bsf}, we see that eq.~\eqref{eq:stuffleExplicit} has a similar basic form but contains a new superscript, also marking its non-abelian nature. 


Equipped with the above rules, we calculate the following fusion product of single-particle generators in an ordering $\alpha$, which are ubiquitous when expanding the commutators like eq.~\eqref{eq:amp}
\begin{align} \label{eq:fusion}
   \widehat \npre(\alpha) &\equiv \ck {\alpha(1)} \star \ck {\alpha(2)} \star  \ldots \star \ck {\alpha({|\alpha|})}\\
   &=t^{\eta}\sum_{r=1}^{|\alpha|-|\eta|}\sum_{\tau\in \mathbf{P}_{\{\tau\}}^{(r)}} (-1)^{|\alpha|-|\eta|-r}  T_{(\tau_1),\ldots,(\tau_r)}^{(\alpha)} \nn \,,
\end{align}
where $t^{\eta}$ is the product of  flavour group generators, and $\mathbf{P}_{\{\tau\}}^{(r)}$ represents all the ordered partitions dividing the gluon ordering $\{\tau\}$ into $r$ sets. The total number of terms is the Fubini number $\mathsf{F}_{|\alpha|-|\eta|}$. Let us consider a simple example  for illustration
\begin{align}
\ck 1  \star \ck 2  & \star  \ck 3  \star\ck 4 =t^{a_1}t^{a_4} \,  T^{(1)} \star T^{(2)}_{(2)} \star T^{(3)}_{(3)} \star T^{(4)} \nonumber \\
 & =t^{a_1}t^{a_4} \left( T^{(1234)}_{(2),(3)}+ T^{(1234)}_{(3),(2)}-T^{(1234)}_{(23)} \right)\,.
\end{align}

The next ingredient of our construction is the map from abstract algebraic generators to functions of physical kinematics and flavour variables 
\begin{align} \label{eq:mapp}
& t^{\eta}T^{(\alpha)}_{(\tau_1),\ldots, (\tau_r )}\,\, {\xrightarrow[\rm{map}]{\langle\bullet \rangle} } \,\,
     \begin{cases}
     	\tr(t^\eta t^{a_n}) \langle T^{(\alpha)}_{(\tau_1),\ldots, (\tau_r )} \rangle_{\A} \, &{\rm amplitude} \\
     ~~	\tr(t^\eta )~~ \langle T^{(\alpha)}_{(\tau_1),\ldots, (\tau_r )} \rangle_{\FF} &{\rm form ~factor}
     \end{cases}
\end{align}
in which $\langle T_{(\tau_1),\ldots,(\tau_r)}^{({\alpha})}\rangle_{\A}$ is defined as follows 
\begin{align}  \label{eq:Talpha}
    \langle & T_{(\tau_1),\ldots,(\tau_r)}^{({\alpha})}  \rangle_{\A}=\Bigg(
    \begin{tikzpicture}[baseline={([yshift=-0.8ex]current bounding box.center)}]\tikzstyle{every node}=[font=\small]    
   \begin{feynman}
    \vertex (a)[myblob]{};
     \vertex[left=0.8cm of a] (a0)[myblobFF]{};
      \vertex[right=0.8cm of a] (a3)[myblob]{};
       \vertex[right=0.8cm of a3] (a4)[myblob]{};
       \vertex[above=0.8cm of a] (b1){$\tau_1$};
        \vertex[above=0.8cm of a3] (b3){$\cdots$};
         \vertex[above=0.8cm of a4] (b4){$\tau_r$};
         \vertex[above=0.8cm of a0] (b0){$ $};
       \vertex [above=0.8cm of a0](j1){$ $};
       \vertex [above=0.5cm of a0](j0F){$ $};
       \vertex [left=1.2cm of j0F](j1F){$~\phi_1~~$};
       \vertex [above=.3cm of j1F](bm1){$\eta$};
       \vertex [below=1.0cm of j1F](j2F){$\phi_{k{-}1}$};
       \vertex [below=0.4cm of j1F](j3F){$\vdots$};
    \vertex [right=0.35cm of j1](j2){$ $};
    \vertex [right=0.8cm of j2](j3){$ $};
      \vertex [right=0.2cm of j3](j6){$ $};
    \vertex [right=0.6cm of j6](j7){$ $};
     \vertex [right=0.0cm of j7](j8){$ $};
    \vertex [right=0.8cm of j8](j9){$ $};
   	 \diagram*{(a)--[very thick](a0),(a)--[very thick](a3), (a3)--[very thick](a4),(a0) -- [thick] (j1F),(a0) -- [thick] (j2F),(a) -- [thick] (j2),(a)--[thick](j3),(a4) -- [thick] (j8),(a4)--[thick](j9)};
    \end{feynman}  
  \end{tikzpicture}\Bigg)^{(\alpha)} \nonumber \\ 
    &={2^r\prod\limits_{i=1}^r \Big(p_{\Theta^{(\alpha)}_L(\tau_i)}\Cdot F_{\tau_i}\Cdot p_{\Theta^{(\alpha)}_R(\tau_i)}\Big) \over  (p_{\eta}^2{-}m^2) (p_{\eta\tau_1}^2{-}m^2)\cdots (p_{\eta\tau_1\cdots \tau_{r-1}}^2{-}m^2)}\, , 
\end{align}
where $p_{X}=\sum_{i\in X} p_{i}$ and $F_{\tau_i}$ represents the product of linearised field strengths  $F^{\mu\nu}_j=p^\mu_j\varepsilon^\nu_j-\varepsilon^{\mu}_j p^{\nu}_j$ for all $j \in \tau_i$. 
Again, the dependence of the $n$-th scalar has been removed via the momentum conservation. Here we have assumed $k>2$. The special case $k=2$ (i.e. the amplitudes with two scalars) is discussed in the Supplemental material, for which the evaluation map requires a minor modification.   For form factors, the  mapping rule is identical, except that $m^2$ in the denominator is replaced by $q^2$, the momentum square of the off-shell operator: $\langle T^{(\alpha)}_{(\tau_1),\ldots, (\tau_r )} \rangle_{\FF}=\langle T^{(\alpha)}_{(\tau_1),\ldots, (\tau_r )} \rangle_{\A}\big |_{m^2\rightarrow q^2}$.

To further clarify $\Theta_{L,R}^{(\alpha)}$, it is illustrative to introduce the ``musical diagram'' as the follow steps. 1. we embed the scalars (denoted as $\eta$) as well as the partitions of gluons $\tau_1$ to $\tau_{r}$ into different levels: $\eta$ lives on the bottom line, $\tau_1$ is above it, then $\tau_2$, until $\tau_r$. 2. we require that when projecting the elements in all the levels onto the bottom line, the ordering should be exactly ${\alpha}$ -- the colour ordering of all the external particles. These requirements uniquely fix the relative positions in both the vertical and the horizontal directions in the ``musical diagram''.  $\Theta^{(\alpha)}_{L,R}(\tau_i)$ are just collections of all the left-lower/right-lower indices of $\tau_i$ in the musical diagram. 
As an example, we consider $T^{({1}567{2}98{3}{4})}_{(578),(69)}$ with the corresponding musical diagram
\begin{align}\label{eq:musicaldiag}
\begin{tikzpicture}[baseline={([yshift=-0.8ex]current bounding box.center)}]\tikzstyle{every node}=[font=\small]    
   \begin{feynman}
    \vertex (l1)[]{}; 
    \vertex [below=0.4cm of l1](ls)[]{}; 
    \vertex [left=0.5cm of ls](rms)[]{$(\eta)$};
    \vertex [right=5.1cm of ls](rs){}; 
    \vertex [left=0.5cm of l1](rm1)[]{$(\tau_1)$};
    \vertex [right=5.1cm of l1](r1)[]{}; 
    \vertex [right=0.5cm of ls](vphif)[myblob2]{\white\small $\mathbf{1}$};
    \vertex [right=2.5cm of ls](vphim)[myblob2]{\white\small $\mathbf{2}$};
     \vertex [right=4.0cm of ls](v3)[myblob2]{\white\small ${\mathbf{3}}$};
      \vertex [right=4.6cm of ls](v4)[myblob2]{\white\small ${\mathbf{4}}$};
    \vertex [right=1.0cm of l1](v5)[sb]{\white\textbf\small $\mathbf{5}$};
    \vertex [right=2.0cm of l1](v7)[sb]{\white\textbf\small $\mathbf{7}$};
    \vertex [right=3.5cm of l1](v8)[sb]{\white\textbf\small $\mathbf{8}$};
    \vertex [above=0.35cm of l1](l2)[]{}; 
    \vertex [right=5.1cm of l2](r2)[]{};
    \vertex [left=0.5cm of l2](rm2)[]{$(\tau_2)$};
    \vertex [right=1.5cm of l2](v6)[sb]{\white\textbf\small $\mathbf{6}$};
    \vertex [right=3.0cm of l2](v9)[sb]{\white\textbf\small $\mathbf{9}$};
   	 \diagram*{(l1)--[thick](v5)- -[thick](v7)- -[thick](v8)--[thick](r1),(ls)--[thick](vphif)--[thick](vphim)--[thick](v3)--[thick](v4)--[thick](rs),(l2)--[thick](v6)--[thick](v9)--[thick](r2)};
    \end{feynman}  
  \end{tikzpicture}
\end{align}
where we denote gluons by discs and scalars with boxes.
Then we have e.g. $p_{\Theta_{L}^{(\alpha)}(\tau_2)}=p_{1}+p_5\equiv p_{15}$ and $p_{\Theta_{R}^{(\alpha)}(\tau_2)}=p_{348}$, so that 
\begin{align}
    &\langle T^{({1}567{2}98{3}{4})}_{(578),(69)}\rangle_{\A} ={4  p_{1}\Cdot F_{578}\Cdot p_{34}\ p_{15}\Cdot F_{69}\Cdot p_{348} \over  (p_{1234}^2{-}m^2) (p_{1234578}^2{-}m^2)} \, .
\end{align}
As a corollary, if  $\Theta^{(\alpha)}_L(\tau_i)$ or $\Theta^{(\alpha)}_R(\tau_i)$ is empty, we then have $p_{\Theta^{(\alpha)}_{L,R}}{=}0$, which leads to the vanishing condition 
\begin{equation}\label{eq:nonvanishing}
    \langle T_{(\tau_1),\ldots,(\tau_r)}^{({\alpha})}\rangle_{\A} = 0\, .
\end{equation}
This is the case if $\alpha$ starts or ends with gluons.

Given these explanations, 
we are now ready to spell out a few examples to illustrate the algebraic construction above. The first example is a four-point amplitude:
\begin{align}\label{eq:fourpteg}
\mathcal{A}(\sc{1},\sc{2},3,\sc{4})={\langle[[\ck 1,\ck 2],\ck 3]\rangle\over p_{12}^2-m^2}+{\langle[\ck 1,[\ck 2,\ck 3]]\rangle\over p_{23}^2 -m^2}\, , 
\end{align}
here and in the following, we denote scalars  by $\bar{i}$. So in the above case, particles $1,2,4$ are scalars and $3$ is a gluon. Expanding the commutators and using the fusion rules together with the mapping rules, we arrive at 
\begin{align}
  \mathcal{A}(\sc{1},\sc{2},3,\sc{4}) &={\langle T^{(231)}_{(3)}\rangle_{\A} \tr([t^{a_1}t^{a_2}]t^{a_4}) \over p_{23}^2-m^2} \cr
&={2p_2\Cdot F_{3}\Cdot p_1 \over (p_{12}^2{-}m^2) (p_{23}^2{-}m^2)} f^{a_1a_2a_4}\, .
  \end{align}
The final expression agrees with the correct amplitude. 
In this example, only the second term in eq.~\eqref{eq:fourpteg} contributes because the BCJ numerator for the first one vanishes as a consequence of eq.~\eqref{eq:nonvanishing}.
The second example is a three-point form factor
 \begin{align}
     \mathcal{F}(\sc{1},3,\sc{2})={\langle[\ck 1,\ck 3]\star\ck 2\rangle\over p_{13}^2-m^2}+{\langle\ck 1\star[\ck 3,\ck 2]\rangle\over p_{23}^2-m^2} \, , 
 \end{align}
which can be simplified to
\begin{align}
\mathcal{F}(\sc 1,  3,\sc 2)
     =  \left({\delta^{a_1 a_2}\over p_{13}^2{-}m^2}+{\delta^{a_1 a_2}\over p_{23}^2{-}m^2}  \right)  {2p_2\Cdot F_{3}\Cdot p_1 \over p_{12}^2{-}q^2 } \, .
\end{align}
The expression agrees with known results \cite{Lin:2021pne}. We have checked our proposal up to all seven-point amplitudes and six-point form factors, as well as eight and night-point ones with two or three scalars. More examples and computation details are given in section C of the Supplemental material and a $\mathsf{Mathematica}$ notebook, which can be found at \cite{ChenGitHub}.
\section{Novel properties of BCJ numerators}
The algebraic construction, in particular the map $\langle \bullet \rangle$, has advantages more than just giving gauge-invariant and duality satisfying numerators. 
Other interesting properties are presented below.


We first start from the following symmetry properties of the map:
\begin{itemize}
\item exchange symmetry: The exchange symmetry for the indices of adjacent scalars   $i,j$
 \begin{align}\label{eq:exchangeSym}
  \langle  T^{(\ldots ij\ldots)}_{(\tau_1), \ldots ,(\tau_r)} \rangle_{\A} =  \langle T^{(\ldots ji\ldots)}_{(\tau_1), \ldots ,(\tau_r)} \rangle_{\A} \, ,
\end{align} 
\item the ``antipode'' symmetry: The antipode symmetry reverses the ordering of particles: 
\begin{align}\label{eq:reverseSym}
 \langle    T_{(\tau_1),\ldots,(\tau_r)}^{({\alpha})} \rangle_{\A} =(-1)^{|\tau|} \langle T_{(\tau_1^{-1}),\ldots,(\tau^{-1}_r)}^{({\alpha}^{-1})} \rangle_{\A} \, ,
\end{align}
where $\alpha^{-1}$  means reversing all the elements in $\alpha$ and the same for $\tau_i^{-1}$, and $|\tau|$ denotes the total number of gluons.
\end{itemize}

Stemming from these symmetries, we have the following three properties of numerators.

(1), we find that the prenumerator, defined as the map of eq.~\eqref{eq:amp}, is invariant under the antipode action 
\begin{align}\label{eq:antipodN}
   &\langle \widehat \npre(12\ldots n{-}1)\rangle\Big|_{t^{a}\rightarrow \Id}= \langle S(\widehat \npre(12\ldots n{-}1))\rangle\Big|_{t^{a}\rightarrow \Id}\, ,
\end{align}
where $S$ is the antipode as an anti-homomorphism  $S(X\star Y)=S(Y)\star S(X)$. The antipode acts on the generators as $S(T^{(i)})=T^{(i)}, S(T^{(j)}_{(j)})=- T^{(j)}_{(j)}$. 
Then \eqref{eq:antipodN} follows from \eqref{eq:reverseSym}.
More details on antipode can be found in \cite{Brandhuber:2022enp} and section B of the Supplemental material.

(2), there is a non-trivial relation between the numerator of the cubic graph corresponding to the left-nested commutator and the corresponding fusion product
\begin{align}
	 \begin{tikzpicture}[baseline={([yshift=-0.0ex,xshift=-0ex]current bounding box.center)}]\tikzstyle{every node}=[font=\small]    \begin{feynman}
    \vertex (a){$\overline{n}$};
     \vertex [above=0.6cm of a](b)[dot]{};
     \vertex [left=0.9cm of b](c);
     \vertex [left=0.22cm of b](c23);
     \vertex [above=0.13cm of c23](v23)[dot]{};
    \vertex [above=.6cm of c](j1){$\alpha(1)$};
    \vertex [right=1.3cm of j1](j2){$\alpha(2)\cdots~~~~$};
    \vertex [right=0.5cm of j2](j3){$~~~~~~~\alpha(n{-}1)$};
   	 \diagram*{(a) -- [thick] (b),(b) -- [thick] (j1),(v23) -- [thick] (j2),(b)--[thick](j3)};
    \end{feynman}  
  \end{tikzpicture}\, &&   \begin{tikzpicture}[baseline={([yshift=-0.0ex,xshift=-0ex]current bounding box.center)}]\tikzstyle{every node}=[font=\small]    \begin{feynman}
    \vertex (a){$\overline{n}$};
     \vertex [above=0.6cm of a](b)[fp]{};
     \vertex [left=0.9cm of b](c);
     \vertex [left=0.22cm of b](c23);
     \vertex [above=0.13cm of c23](v23)[fp]{};
    \vertex [above=.6cm of c](j1){$\alpha(1)$};
    \vertex [right=1.3cm of j1](j2){$\alpha(2)\cdots~~~~$};
    \vertex [right=0.5cm of j2](j3){$~~~~~~~\alpha(n{-}1)$};
   	 \diagram*{(a) -- [thick] (b),(b) -- [thick](v23) -- [thick] (j1),(v23) -- [thick] (j2),(b)--[thick](j3)};
    \end{feynman}  
  \end{tikzpicture}\nn\\
\langle\widehat\npre([\alpha])\rangle~~~~~~~~~~~ && \langle\widehat\npre(\alpha)\rangle~~~~~~~~~~~
\end{align}
which is known as the BCJ numerator and pre-numerator respectively \footnote{Recall, the black circle vertices represent nested commutators, whereas the red box vertices correspond to fusion products.}. The flavour factors of the two numerators are $\tr([t^{\eta}]t^{a_n}) $  and $\tr(t^{\eta}t^{a_n})$ respectively, where $t^{\eta}$ denotes the product of the flavour generators $t^{a_i}$ for $i \in \eta$, and $[t^{\eta}]$ represents the nest commutator of these generators.  Magically, the kinematic part of the BCJ numerator and the pre-numerator are identical
\begin{align}\label{eq:preNum2DDMNum}
 \langle\widehat\npre([\alpha])\rangle\big|_{f^{a b c}\rightarrow1}&= \langle\widehat\npre(\alpha)\rangle\big|_{t^{a}\rightarrow\Id} \,.
\end{align}
A simple example is 
\begin{align}
\langle\npre(\sc 1,\sc 2, 3,\sc 4)\rangle &=\langle  \ck 1\star \ck 2\star \ck 3 \star \ck 4 \rangle \nn\\
&={2p_{12}\Cdot F_3\Cdot p_4 \over p^2_{124}-m^2}\tr(t^{a_1}t^{a_2}t^{a_4}t^{a_5})\, , \\
	\langle\npre([\sc 1,\sc 2,3,\sc 4])\rangle &=\langle  [\ck 1, \ck 2]\star \ck 3 \star \ck 4+\ck 4\star \ck 3\star [\ck 1, \ck 2] \rangle \nn\\
	&={2p_{12}\Cdot F_3\Cdot p_4 \over p^2_{124}-m^2}\tr([[t^{a_1},t^{a_2}],t^{a_4}]t^{a_5}) \, .
\end{align}

(3), for form factors of ${\rm Tr}{(\phi^2)}$, another relation arises\footnote{Such relations are actually requirements from having a consistent double copy, see \cite{Lin:2021pne,Lin:2022jrp}}
\begin{align} \label{eq:ff-relation}
 \langle\widehat\npre([\Gamma_1,j])\star \widehat\npre(\Gamma_2)\rangle &= \langle\widehat\npre(\Gamma_1)\star \widehat\npre([j,\Gamma_2])\rangle \nn\\
\begin{tikzpicture}[baseline={([yshift=-0.8ex]current bounding box.center)}]\tikzstyle{every node}=[font=\small]    
   \begin{feynman}
    \vertex (a)[]{$q$};
     \vertex [above=0.6cm of a](b)[fp]{};
     \vertex [left=1.cm of b](c);
     \vertex [left=0.3cm of b](c1);
     \vertex [left=0.6cm of b](cL);
     \vertex [above=0.4cm of c1](b1)[dot]{};
     \vertex [above=0.63cm of cL](vL)[HV]{\tiny$\Gamma_{1}$};
     \vertex [right=1.2cm of vL](vR)[HV]{\tiny$\Gamma_{2}$};
     \vertex [above=0.6cm of vL](t1){$\sigma_{1}~$};
     \vertex [above=0.6cm of vR](t2){$~~\sigma_{2}$};
    \vertex [above=1.4cm of c](j1){};
    \vertex [right=.7cm of j1](j2){};
     \vertex [right=0.45cm of j2](j5){$j$};
      \vertex [right=0.25cm of j5](j6){};
       \vertex [right=0.6cm of j6](j7){};
   	 \diagram*{(a) -- [blue, ultra thick] (b),(b)--[thick](vL) -- [thick] (j1),(vL) -- [thick] (j2),(b1)--[thick](j5), (vR)--[thick](j6),(b)--[thick](vR)--[thick](j7)};
    \end{feynman}  
  \end{tikzpicture}&~~~~~~
  \begin{tikzpicture}[baseline={([yshift=-0.8ex]current bounding box.center)}]\tikzstyle{every node}=[font=\small]    
   \begin{feynman}
    \vertex (a)[]{$q$};
     \vertex [above=0.6cm of a](b)[fp]{};
     \vertex [left=1.cm of b](c);
     \vertex [right=0.3cm of b](c1);
     \vertex [left=0.6cm of b](cL);
     \vertex [above=0.4cm of c1](b1)[dot]{};
     \vertex [above=0.63cm of cL](vL)[HV]{\tiny$\Gamma_{1}$};
     \vertex [right=1.2cm of vL](vR)[HV]{\tiny$\Gamma_{2}$};
     \vertex [above=0.6cm of vL](t1){$\sigma_{1}~$};
     \vertex [above=0.6cm of vR](t2){$~~\sigma_{2}$};
    \vertex [above=1.4cm of c](j1){};
    \vertex [right=.7cm of j1](j2){};
     \vertex [right=0.3cm of j2](j5){$j$};
      \vertex [right=0.4cm of j5](j6){};
       \vertex [right=0.6cm of j6](j7){};
   	 \diagram*{(a) -- [blue, ultra thick] (b),(b)--[thick](vL) -- [thick] (j1),(vL) -- [thick] (j2),(b1)--[thick](j5), (vR)--[thick](j6),(b)--[thick](vR)--[thick](j7)};
    \end{feynman}  
  \end{tikzpicture} .
\end{align}
Here the $j$ line can be either gluon or scalar. When $j$ line is gluon, the identity is manifest according to the vanishing condition eq.~\eqref{eq:nonvanishing}. 
 If $j$ line is scalar, the identity becomes highly non-trivial, e.g.  $\langle\widehat\npre([\sc 1, 2])\star\widehat\npre([\sc 3,\sc 4])-\widehat\npre([[\sc 1, 2],\sc 3])\star \widehat\npre(\sc 4) \rangle$ is evaluated as
  \begin{align}
 	&{-}\langle \ck 1\star \ck 2\star \ck 4\star \ck 3\rangle{+}\langle \ck 3\star [\ck 1, \ck 2]\star \ck 4\rangle= \tr(t^{a_1}t^{a_4}t^{a_3}) \nn\\
 	&\Big(-{p_{1}\Cdot F_2\Cdot p_{34}\over p^2_{134}-q^2}+{p_{13}\Cdot F_2\Cdot p_{4}\over p^2_{134}-q^2}-{p_{3}\Cdot F_2\Cdot p_{14}\over p^2_{134}-q^2}\Big)=0 \, ,
 \end{align}
 which implies,
  \begin{align}
 \langle\widehat\npre([\sc 1, 2])\star\widehat\npre([\sc 3,\sc 4]) \rangle = \langle \widehat\npre([[\sc 1, 2],\sc 3])\star \widehat\npre(\sc 4) \rangle \, .
 \end{align}

\section{Conclusions and Discussions}

In this letter, we proposed a kinematic algebra for the BCJ numerators in YMS$+ \phi^3$ theory. 
The underlying algebraic structures lead to extremely compact expressions for the BCJ numerators both in amplitudes and form factors, and reveal intriguing relations among them.
Besides manifestly obeying the Jacobi identities, the numerators constructed in this way also enjoy many other remarkable properties such as crossing symmetry, manifest gauge invariance and antipode symmetry. 


The amplitudes and BCJ numerator in the YMS$+ \phi^3$ theory has important application to constructing the gravitational amplitudes via double copy and studying the gravitational physics. 
For example, when double-copied with pure YM amplitudes, Einstein-Yang-Mills and Einstein-Maxwell amplitudes can be obtained, which are useful  in the gravitational scattering of photon from black hole \cite{Bjerrum-Bohr:2014lea,Chen:2022yxw,Kim:2022iub}.
Moreover, when double-copied with amplitudes of (massive) spinning particles coupled to gluons, the resulting amplitudes are involved in the study of  black hole scattering with spin effects \footnote{If we are only interested in the spinless black holes \cite{Kosower:2018adc,Bern:2019nnu,Damour:2019lcq,DiVecchia:2021bdo,Herrmann:2021tct,Bjerrum-Bohr:2021din,Bern:2021dqo, Bjerrum-Bohr:2021wwt,Jakobsen:2021lvp}, the HEFT amplitudes is sufficient \cite{Brandhuber:2021eyq} because their double copy fully capture the classical piece of scalar-graviton scattering. However, involving spinning particles requires more, and the Yang-Mills-scalar numerators/amplitudes obtained in this paper are needed.}, see \cite{Guevara:2017csg,Vines:2018gqi, Guevara:2018wpp, Chung:2018kqs, Arkani-Hamed:2019ymq, Guevara:2019fsj, Chung:2019duq, Damgaard:2019lfh, Aoude:2020onz, Chung:2020rrz, Guevara:2020xjx,Bern:2020buy, Kosmopoulos:2021zoq,Chen:2021qkk,Bern:2022kto,Aoude:2022trd, Aoude:2022thd}.


One more application is as follows. 
The colour-kinematic duality and double copy have also been studied in some effective theories with higher-dimensional interactions \cite{Elvang:2018dco,Carrasco:2019yyn,Carrasco:2021ptp,Chi:2021mio,Bonnefoy:2021qgu}. As a step towards such direction, one may consider form factors with the insertion of higher-dimensional operators. In the Supplemental material, we show that the algebraic construction also works directly for form factors with such operators. Interestingly, novel relations beyond Jacobi identity are deduced naturally from the  kinematic algebra. 

We now give some outlooks. First, at tree-level, one can extend the applicable scope of the Hopf algebra to more general theories; giving a proof also deserves considerations. Second, a feasible direction is to explore the kinematic algebra at the level of loop integrands. The physical picture of the fusion products (especially when involving the internal lines) suggests that they can be readily generalised to off-shell particles. Third, it would be fascinating to find connections between our construction and other approaches in the literature, such as the Lagrangian and geometric understanding of the colour-kinematics duality \cite{Fu:2016plh,Cheung:2016prv,Cheung:2017yef, Reiterer:2019dys,Tolotti:2013caa,Mizera:2019blq,Borsten:2020zgj,Borsten:2020xbt,Ferrero:2020vww,Borsten:2021hua,Cheung:2021zvb,Cheung:2022vnd,Cohen:2022uuw,Diaz-Jaramillo:2021wtl,Guevara:2021yud,White:2020sfn}, especially regarding the close relation between quasi-shuffle algebra and the permutohedron geometry \cite{Cao:2021dcd,Cao:2022vou} (see also \cite{kapranov1993permutoassociahedron,billera1994combinatorics,Tonks95relatingthe,postnikov2005math7163P,postnikov2006faces}).

\section*{Acknowledgements}
We thank Andreas Brandhuber, Graham Brown, Josh Gowdy, Gabriele Travaglini for useful discussions and collaboration on related topics. We would also like to thank Song He, Marco Chiodaroli, Oliver Schlotterer, Gang Yang and Mao Zeng for insightful discussions. This work was supported by the Science and Technology Facilities Council (STFC) Consolidated Grants ST/P000754/1 \textit{``String theory, gauge theory \& duality''},  ST/T000686/1 \textit{``Amplitudes, strings  \& duality''} and by the European Union’s Horizon 2020 research and innovation programme under the Marie Skłodowska-Curie grant agreement No. 764850 “SAGEX”. 
GL and CW are supported by a Royal Society University Research Fellowship No.~UF160350. GL also thanks the Higgs Centre for Theoretical Physics at the University of Edinburgh for hospitality. No new data were generated or analysed during this study.

\newpage
\appendix
\onecolumngrid

\section{\textbf{A.} A Special mapping rule for the amplitudes with two scalars }\label{app:twoscalar}

As we commented in the main text, even though the BCJ numerators of amplitudes with two scalars and arbitrary number of gluons can be constructed from the same kinematic algebra, the evaluation map requires a slight modification. 
These amplitudes are special because there is only one independent scalar momentum $p_{\phi}$ (the other one $\phi^{\prime}$, set to be $\overline{n}$ as usual, is eliminated by the momentum conservation). 
If the vanishing condition eq.~\eqref{eq:nonvanishing} in the main text were not modified, all the numerators would be zero. 
To circumvent this difficulty, in the two-scalar case, we treat one gluon together as if it behaves like a scalar. 
For convenience, we set it to be $1^{g}$ and we also have the other scalar $\sc{n{-}1}$. 
They are denoted as $\eta$ together lying on the bottom line of the ``musical diagrams''. 

The BCJ numerator with two scalars from the kinematic algebra is given by 
\begin{align} \label{eq:star}
\npre(\alpha, \sc n)=\langle\ck{\alpha(1)}\star \ck{\alpha(2)}\star\cdots\star \ck{\alpha(n-1)}\rangle=\tr(t^{a_{n-1}}t^{a_n}) \sum_{r=1}^{n{-}2}\sum_{\tau\in \mathbf{P}_{\{\tau\}}^{(r)}} (-1)^{n-r} \sum_{i=1}^{r} \langle T^{(\alpha)}_{(\tau_1),\ldots,(\tau_{i-1}),(\widetilde\tau_i),(\tau_{i+1}),(\tau_r)}\rangle_{\A} \, ,
\end{align}
where $\tilde\tau_i$ is the set that contains the gluon $1$. First, analogous to eq.~\eqref{eq:Talpha}, 
we have
\begin{align} \label{eq:000}
    \langle T^{(\alpha)}_{(\tau_1),\ldots,(\tau_r)}\rangle_{\A}=0 \, ,
\end{align}
unless $\alpha$ begins and ends with the gluon $1$ and scalar $\sc{n{-}1}$, and gluon $1$ must be with $\tau_1$. 
These two requirements further imply that we can only have $(1\tau_1)$ or $(\tau_1 1)$ in the subscript of $T$. 
Therefore, only two types of generators are non-vanishing after the evaluation map: \textbf{(a):} $\langle T^{(1 \alpha' {n{-}1})}_{(1\tau_1),(\tau_2)\ldots,(\tau_r)}\rangle_{\A}$, and \textbf{(b):} $\langle T^{(n{-}1 \,\alpha' 1)}_{(\tau_11),\ldots,(\tau_r)}\rangle_\A$ with
\begin{equation}\label{eq:2scalarantipod}
   \langle T^{(n{-}1 \,\alpha' 1)}_{(\tau_11),\ldots,(\tau_r)}\rangle_\A = (-1)^{n{-}2}\langle T^{(1\alpha^{\prime-1} \,n-1)}_{(1\tau^{-1}_1 ),\ldots,(\tau^{-1}_r)}\rangle_\A \, ,
\end{equation}
which is related to \textbf{(a)} via the antipode action from eq.~\eqref{eq:reverseSym}. 
The corresponding musical diagrams are given by
\begin{align}\label{eq:musicNonzero}
\begin{tikzpicture}[baseline={([yshift=-0.8ex]current bounding box.center)}]\tikzstyle{every node}=[font=\small]    
   \begin{feynman}
    \vertex (l1)[]{}; 
    \vertex [below=0.4cm of l1](ls)[]{}; 
    \vertex [above=0.6cm of l1](lg)[]{$\vdots~~~~~~~~~$}; 
     \vertex [right=1.0cm of lg](lg3)[]{$\vdots$}; 
    \vertex [right=3.5cm of lg](lg2)[]{$\vdots$}; 
    \vertex [left=0.5cm of ls](rms)[]{$(\eta)$};
    \vertex [right=5.cm of ls](rs){}; 
    \vertex [right=4.3cm of ls](vphi)[myblob2]{\white\textbf\small $\mathbf{n{-}1}$}; 
    \vertex [left=0.5cm of l1](rm1)[]{$(\tau_1)$};
    \vertex [right=5.cm of l1](r1)[]{}; 
    \vertex [right=0.5cm of ls](v1)[sb]{\white\textbf\small $\mathbf{1}$};
    \vertex [right=1.cm of l1](v3)[sb]{\white\textbf\small $\mathbf{~\,~}$};
    \vertex [right=3.5cm of l1](v4)[sb]{\white\textbf\small $\mathbf{~\,~}$};
   	 \diagram*{(l1)- -[thick](v3)- -[scalar,thick](v4)--[thick](r1),(ls)--[thick](v1)--[thick](vphi)--[thick](rs)};
    \end{feynman}  
  \end{tikzpicture} && 
  \begin{tikzpicture}[baseline={([yshift=-0.8ex]current bounding box.center)}]\tikzstyle{every node}=[font=\small]    
   \begin{feynman}
    \vertex (l1)[]{}; 
    \vertex [below=0.4cm of l1](ls)[]{}; 
    \vertex [above=0.6cm of l1](lg)[]{$\vdots~~~~~~~~~$}; 
     \vertex [right=1.5cm of lg](lg3)[]{$\vdots$}; 
    \vertex [right=4.0cm of lg](lg2)[]{$\vdots$}; 
    \vertex [left=0.5cm of ls](rms)[]{$(\eta)$};
    \vertex [right=5.cm of ls](rs){}; 
    \vertex [right=4.5cm of ls](vphi)[sb]{\white\textbf\small $\mathbf{1}$}; 
    \vertex [left=0.5cm of l1](rm1)[]{$(\tau_1)$};
    \vertex [right=5.cm of l1](r1)[]{}; 
    \vertex [right=0.7cm of ls](v1)[myblob2]{\white\textbf\small $\mathbf{n{-}1}$};
    \vertex [right=1.5cm of l1](v3)[sb]{\white\textbf\small $\mathbf{~\,~}$};
    \vertex [right=4.0cm of l1](v4)[sb]{\white\textbf\small $\mathbf{~\,~}$};
   	 \diagram*{(l1)- -[thick](v3)- -[scalar,thick](v4)--[thick](r1),(ls)--[thick](v1)--[thick](vphi)--[thick](rs)};
    \end{feynman}  
  \end{tikzpicture}\, .
\end{align}
For example, for a ten-point case, $\langle T^{(132756489)}_{(127),(68),(4),(35)}\rangle_{\A}$ (once again we have removed the $10$-th scalar), the musical diagram is given as 
\begin{align}\label{eq:musicaldiag2}
\begin{tikzpicture}[baseline={([yshift=-0.8ex]current bounding box.center)}]\tikzstyle{every node}=[font=\small]    
   \begin{feynman}
    \vertex (l1)[]{}; 
    \vertex [below=0.4cm of l1](ls)[]{}; 
    \vertex [left=0.5cm of ls](rms)[]{$(\eta)$};
    \vertex [right=5.cm of ls](rs){}; 
    \vertex [right=4.5cm of ls](vphi)[myblob2]{\white\textbf\small $\mathbf{9}$}; 
    \vertex [left=0.5cm of l1](rm1)[]{$(\tau_1)$};
    \vertex [right=5.cm of l1](r1)[]{}; 
    \vertex [right=0.5cm of ls](v1)[sb]{\white\textbf\small $\mathbf{1}$};
    \vertex [right=1.5cm of l1](v3)[sb]{\white\textbf\small $\mathbf{2}$};
    \vertex [right=2.0cm of l1](v4)[sb]{\white\textbf\small $\mathbf{7}$};
    \vertex [above=0.3cm of l1](l2)[]{};  
    \vertex [right=5.cm of l2](r2)[]{};
    \vertex [left=0.5cm of l2](rm2)[]{$(\tau_2)$};
    \vertex [right=3.0cm of l2](v6)[sb]{\white\textbf\small $\mathbf{6}$};
     \vertex [right=4.cm of l2](v9)[sb]{\white\textbf\small $\mathbf{8}$};
     \vertex [above=0.3cm of l2](l3)[]{};  
     \vertex [right=5.cm of l3](r3)[]{};
     \vertex [left=0.5cm of l3](rm3)[]{$(\tau_3)$};
     \vertex [right=3.5cm of l3](v7)[sb]{\white\textbf\small $\mathbf{4}$};
     \vertex [above=0.3cm of l3](l4)[]{};
     \vertex [right=5.cm of l4](r4)[]{};
     \vertex [left=0.5cm of l4](rm4)[]{$(\tau_4)$};
     \vertex [right=1.cm of l4](v2)[sb]{\white\textbf\small $\mathbf{3}$};
      \vertex [right=2.5cm of l4](v5)[sb]{\white\textbf\small $\mathbf{5}$};
   	 \diagram*{(l1)- -[thick](v3)- -[thick](v4)--[thick](r1),(l2)--[thick](v6)--[thick](v9)--[thick](r2),(l3)--[thick](v7)--[thick](r3),(l4)--[thick](v2)--[thick](v5)--[thick](r4),(ls)--[thick](v1)--[thick](vphi)--[thick](rs)};
    \end{feynman}  
  \end{tikzpicture}\, .
\end{align}
The mapping rule for the multi-scalar cases in the main text applies identically, except for $F_{\tau_1}$, because we must include $F_1$ in the final result. Inspired by the mapping rules for the BCJ numerators with two heavy particles \cite{Brandhuber:2021bsf}, we propose
\begin{align}\label{eq:T}
	&\langle T^{(1\alpha^{\prime}{{n{-}1}})}_{(1\tau_1),\ldots,(\tau_r)}\rangle_\A= \Bigg(\begin{tikzpicture}[baseline={([yshift=-0.8ex]current bounding box.center)}]\tikzstyle{every node}=[font=\small]    
   \begin{feynman}
    \vertex (a)[myblob]{};
     \vertex[left=0.8cm of a] (a0)[myblobFF]{};
      \vertex[right=0.8cm of a] (a3)[myblob]{};
       \vertex[right=0.8cm of a3] (a4)[myblob]{};
       \vertex[above=0.8cm of a] (b1){$\tau_1$};
        \vertex[above=0.8cm of a3] (b3){$\cdots$};
         \vertex[above=0.8cm of a4] (b4){$\tau_r$};
         \vertex[above=0.8cm of a0] (b0){$ $};
       \vertex [above=0.8cm of a0](j1){$ $};
       \vertex [above=0.5cm of a0](j0F){$ $};
       \vertex [left=1.0cm of j0F](j1F){$~~1~~$};
       \vertex [above=.3cm of j1F](bm1){$\eta$};
       \vertex [below=1.0cm of j1F](j2F){${n{-}1}$};
       \vertex [below=0.4cm of j1F](j3F){$~$};
    \vertex [right=0.35cm of j1](j2){$ $};
    \vertex [right=0.8cm of j2](j3){$ $};
      \vertex [right=0.2cm of j3](j6){$ $};
    \vertex [right=0.6cm of j6](j7){$ $};
     \vertex [right=0.0cm of j7](j8){$ $};
    \vertex [right=0.8cm of j8](j9){$ $};
   	 \diagram*{(a)--[very thick](a0),(a)--[very thick](a3), (a3)--[very thick](a4),(a0) -- [thick] (j1F),(a0) -- [thick] (j2F),(a) -- [thick] (j2),(a)--[thick](j3),(a4) -- [thick] (j8),(a4)--[thick](j9)};
    \end{feynman}  
  \end{tikzpicture}\Bigg)^{(1\alpha^{\prime}{n{-}1} )}
=-{2^{r} p_{n{-}1}\Cdot F_{1}\Cdot F_{\tau_1}\Cdot p_{n{-}1}\prod\limits_{i=2}^r \Big(p_{\Theta^{(\alpha)}_L(\tau_i)}\Cdot F_{\tau_i}\Cdot p_{\Theta^{(\alpha)}_R(\tau_i)}\Big)\over 2\, (p_{\eta}^2{-}m^2) (p_{\eta\tau_1}^2{-}m^2)\cdots (p_{\eta\tau_1\cdots \tau_{r{-}1}}^2{-}m^2)}\, ,
\end{align}
where $(\alpha)$ is $(1\alpha^{\prime}{{n{-}1}})$,  $p_{{\eta}} = p_1 + p_{n-1}$ and $p_{n-1}$ is the momentum of the remaining scalar. 
As we commented earlier, the terms involving $\tau_2$ to $\tau_{r}$ are obtained following exactly the same rules as the multi-scalar cases.  
Furthermore, the expression reproduces exactly that for HEFT in  \cite{Brandhuber:2021bsf} in the heavy-mass limit, which amounts to the replacement: $p_{n-1}{+}p_{i_1\ldots i_r}\rightarrow m v$ (therefore $(p_{n-1}{+}p_{i_1\ldots i_r})^2-m^2\rightarrow 2mv\Cdot p_{i_1\ldots i_r}$). As an example, for the musical diagram in eq.~\eqref{eq:musicaldiag2}, we have 
\begin{align}
    & \langle T^{(132756489)}_{(127),(68),(4),(35)}\rangle_\A=- {8 (p_{9}\Cdot F_1\Cdot F_{27}\Cdot p_{9})(p_{134}\Cdot F_{68}\Cdot p_{9})(p_{1346}\Cdot F_{4}\Cdot p_{9})(p_{1}\Cdot F_{35}\Cdot p_{6489}) \over (p_{ 19}^2-m^2)(p_{1279}^2-m^2)(p_{126789}^2-m^2)(p_{1246789}^2-m^2)} \, .
\end{align}
Note that the above definition of numerators cannot apply to the three-point case, for which the map is given by
\begin{align}
   \npre(1,\sc 2, \sc 3)=\langle \ck 1\star \ck2\rangle=\langle T_{(1)}^{(12)}\rangle_{\A}=-{1\over 2}p_2\Cdot \varepsilon_1 \,,&&
    \npre(\sc 2,1,\sc 3)=\langle \ck 2\star \ck 1\rangle=\langle T_{(1)}^{(21)}\rangle_{\A}={1\over 2}p_2\Cdot \varepsilon_1\,.
\end{align}
We have checked up to nine points that these numerators produce the correct amplitudes. 

\section{\textbf{B.} A Minimal introduction to the non-abelian quasi-shuffle Hopf algebra}\label{sec:hofpAlgebra}
 
 For the algebraic generators to form a quasi-shuffle Hopf algebra, they should have compatible coproduct, counit and antipode, which are denoted as $\cop, \epsilon$ and $S$, respectively. In this section, we will give a minimal introduction to these mathematical structures.  The Hopf algebra of the non-abelian quasi-shuffle product generators that are relevant for amplitudes is developed in \cite{Brandhuber:2022enp} and the actions of the coproduct,  counit and antipode on the generators are given by
 \begin{equation}
\begin{aligned}\label{eq: extendedcoproduct}
\begin{split}
\cop(T^{(\alpha)})&=\left(T^{(\alpha)}\otimes T^{(\alpha)}\right)\,, \\
    \cop(T_{(\tau_1),\ldots, (\tau_r )}^{(\alpha)})&=\sum_{i=0}^{r}\left(T_{(\tau_1),\ldots, (\tau_i )}^{(\alpha)}\otimes T_{(\tau_{i+1}),\ldots, (\tau_r )}^{(\alpha)}\right)\,, \\
    \epsilon(T^{(\alpha)}_{\tau})&= 0\, , \qquad \qquad \quad
     \epsilon(T^{(\alpha)})= \Id_{\rm qs}\,\\
     S(T^{(\alpha)}_{(\tau_1),\ldots, (\tau_r)})
&=\sum_{h=0}^{r-1}(-1)^{h+1}\sum_{r>i_1>i_2\cdots>i_h>1}T^{(\alpha^{-1})}_{(\tau_{r}\cdots\tau_{i_1+1}),(\tau_{i_1}\cdots\tau_{i_2+1}),\cdots,(\tau_{i_h}\cdots\tau_{1})}\, , \\
S(T^{(\alpha)})&=T^{(\alpha^{-1})}\, ,
    \end{split} 
\end{aligned}
\end{equation}
where $T^{(\alpha^{-1})}\star T^{(\alpha)}= T^{(\alpha)}\star T^{(\alpha^{-1})}=\Id_{\rm qs}$ and $\Id_{\rm qs}$ is the identity element in the algebra with the property $\Id_{\rm qs}\star T^{(\alpha)}_\tau=T^{(\alpha)}_\tau\star \Id_{\rm qs}=T^{(\alpha)}_\tau$. 
The compatible conditions are 
\begin{equation}
\begin{aligned}  
\Delta(T^{(\alpha)}_{\tau})\star \Delta(T^{(\beta)}_{\omega})&=\Delta(T^{(\alpha)}_{\tau}\star T^{(\beta)}_{\omega})\,,\\
\epsilon(T^{(\alpha)}_{\tau})\star \epsilon(T^{(\beta)}_{\omega})&=\epsilon(T^{(\alpha)}_{\tau}\star T^{(\beta)}_{\omega}) \,,\\
\star(\Id_{\rm qs}\otimes S)\Delta(T^{(\alpha)}_{\tau})&=\star( S\otimes \Id_{\rm qs})\Delta(T^{(\alpha)}_{\tau})=\epsilon(T^{(\alpha)}_{\tau})\Id_{\rm qs} \,,\\
S(T^{(\alpha)}_{\tau}\star T^{(\beta)}_{\omega})&=S(T^{(\beta)}_{\omega})\star S(T^{(\alpha)}_{\omega})\, .
\end{aligned}
\end{equation}

As discussed in the main text, the numerators that are related by the antipode obey interesting relations. Here we take a five-point numerator with two gluons as an example. After omitting the flavor factors, it is given as
\begin{align}\label{eq:S1234}
   \widehat\npre(\sc 1, 3,4,\sc 2)|_{t^a\rightarrow \Id}=  T^{(1)}\star T^{(3)}_{(3)}\star T^{(4)}_{(4)}\star T^{(2)}\, .
\end{align}
The action of the antipode gives 
\begin{align}\label{eq:S4321}
  S( \widehat\npre(\sc 1, 3,4,\sc 2)|_{t^a\rightarrow \Id})= S( T^{(1)}\star T^{(3)}_{(3)}\star T^{(4)}_{(4)}\star T^{(2)})&=S(T^{(2)})\star S(T^{(4)}_{(4)})\star S(T^{(3)}_{(3)}) \star S(T^{(1)})\nn\\
  &=(-1)^2 T^{(2)}\star T^{(4)}_{(4)}\star T^{(3)}_{(3)} \star T^{(1)}\, .
\end{align}
Apply the map rules on eqs.~\eqref{eq:S1234} and \eqref{eq:S4321}, we get
\begin{align}
 \langle T^{(1)}\star T^{(3)}_{(3)}\star T^{(4)}_{(4)}\star T^{(2)}\rangle_{\A,\FF} &=\langle -T^{(1342)}_{(34)}+T^{(1342)}_{(3),(4)}+T^{(1342)}_{(4),(3)}\rangle_{\A,\FF}\,,\nn\\
   \langle (-1)^2 T^{(2)}\star T^{(4)}_{(4)}\star T^{(3)}_{(3)} \star T^{(1)}\rangle_{\A,\FF} &=\langle -T^{(2431)}_{(43)}+T^{(2431)}_{(3),(4)}+T^{(2431)}_{(4),(3)}\rangle_{\A,\FF}\, .
\end{align}
Using the antipode relation eq.~\eqref{eq:reverseSym} in the main text, 
it is easy to see that the right hand side of the above two equations are equivalent. Hence the pre-numerator is invariant under the antipode (when ignoring the flavour factors)
\begin{align}
   \langle \widehat\npre(\sc 1, 3,4,\sc 2)|_{t^a\rightarrow \Id}\rangle=\langle S( \widehat\npre(\sc 1, 3,4,\sc 2)|_{t^a\rightarrow \Id})\rangle \,.
\end{align}

\section{\textbf{C.} More examples on the Kinematic Hopf algebra and BCJ numerators}

In this section, we consider more concrete examples to illustrate the Kinematic Hopf algebra construction of amplitude. In practice, such construction implies another closed form of amplitude and form factor by   introducing the pre-numerators and propagator matrix. Their pre-numerators $\npre(\alpha,\widebar n),\npre_\mathcal{F}(\alpha,\widebar n) $ are defined as performing the evaluation map on the fusion product $ \widehat\npre(\alpha)$.
Given the pre-numerators, we can directly obtain the double ordered amplitudes as
\begin{align}\label{eq:numtoA}
     \mathcal{A}&(\eta, \overline{n}|\sigma, \overline{n})\tr(t^{\eta} t^{a_{n}})\\
   \nonumber &=\hskip -6pt \sum_{\substack{\alpha^{\prime} \in  S_{n{-}3}/S_{k{-}3}}}\hskip -6pt
    \mathsf{m}(\sigma\overline{n}|\sc 1 \alpha^{\prime} \sc {k{-}1} \overline{n}) \npre(\sc 1, \alpha^{\prime}, \sc {k{-}1},\overline{n})\,,
\end{align}
where $\eta=\{\sc 1\ldots \sc{k{-}1}\}$  represents the ordered scalars, with $t^{\eta}$ being the product of flavour generators of the scalars in $\eta$. The sum is over permutations preserving the relative ordering of $(k{-}3)$ scalars $\sc 2,\ldots,\sc {k{-}2}$. Moreover, $\mathsf{m}$ is the propagator matrix \cite{Vaman:2010ez} encoding the bi-adjoint $\phi^3$ amplitudes \cite{Du:2011js,BjerrumBohr:2012mg,Cachazo:2013iea,Arkani-Hamed:2017mur,Bahjat-Abbas:2018vgo}. Summing over all flavour traces, we have
\begin{equation}
\begin{aligned}
      \mathcal{A}(\sigma, \overline{n})&=\sum_{\beta \in S_{k{-}1}}\mathcal{A}(\beta(\eta),\overline{n}|\sigma,\overline{n})\tr(t^{\beta(\eta)} t^{a_n})\\
      &=\sum_{\alpha\in S_{n{-}1}}
    \mathsf{m}(\sigma\overline{n}|\alpha \overline{n}) \npre(\alpha,\overline{n})\,,
\end{aligned} \label{eq:Amatrix}
\end{equation}
where $\beta$ permutes scalars only, $\alpha$ acts on all particles except $\overline{n}$ and the vanishing condition eq.~\eqref{eq:nonvanishing}  is employed. Similarly, for an $n$-point form factor, we have 
\begin{equation}
    \mathcal{F}(\sigma)=\sum_{\gamma\in S_{n}/\mathbb{Z}_{n}}\mathsf{m}_{\cal F}(\sigma|\alpha)\npre_{\cal F}(\alpha)\,.
\end{equation}
The fact that the operator is a colour/flavour singlet not only affects $\npre_{\cal F}$ but also alters the propagator matrix $\mathsf{m}_{\cal F}(\sigma|\alpha)$ for form factors. This matrix can be obtained by summing up all possible operator insertions in the cubic diagrams involved in $\mathsf{m}(\sigma|\alpha)$ \cite{Lin:2021pne}.

We begin by considering examples for the amplitudes with two scalars. For a four-point amplitude $A(1,2,\sc 3, \sc 4)$, the only non-vanishing pre-numerators are given in eq.~\eqref{eq:T}
\begin{align}
   \npre(1,2,\sc 3, \sc 4)&=\langle \ck 1\star \ck2\star \ck3\rangle=\langle T_{(1),(2)}^{(123)}+T_{(2),(1)}^{(123)}-T_{(12)}^{(123)}\rangle_{\A} \, \tr(t^{a_3}t^{a_4}) ={p_3\Cdot F_1\Cdot F_2\Cdot p_3\over p_{13}^2-m^2} \tr(t^{a_3}t^{a_4})\, ,
\end{align}
and  $\npre(\sc 3, 2,1,\sc 4)= \npre(1,2,\sc 3, \sc 4)$. Then the amplitude is obtained directly from  eq.~\eqref{eq:numtoA}
\begin{equation} \label{eq:a1234}
    \mathcal{A}(1,2,\sc 3,\sc 4)= 2\,\mathsf{m}(1234|1234)\mathcal{N}(1,2,\sc 3,\sc 4)=2\bigg(\frac{1}{p_{12}^2 }+\frac{1}{p_{23}^2-m^2}\bigg)\mathcal{N}(1,2,\sc 3,\sc 4)\,.
\end{equation}
The factor $2$ in the above equation is due to the antipode symmetry in eq.~\eqref{eq:2scalarantipod} and the fact that we have a special flavour factor in the two scalar case, $\operatorname{tr}(t^{a_3}t^{a_4})=\operatorname{tr}(t^{a_4}t^{a_3})$. We will omit this simple flavour factor below. 
In terms of the nested commutators, the four-point amplitude is given by 
\begin{align}\label{eq:A4ck}
{\cal A}(1,2,\sc{3},\sc 4)&=
 \begin{tikzpicture}[baseline={([yshift=-0.8ex]current bounding box.center)}]\tikzstyle{every node}=[font=\small]    
   \begin{feynman}
    \vertex (a){$\sc 4$};
     \vertex [above=0.7cm of a](b)[dot]{};
     \vertex [left=0.7cm of b](c);
     \vertex [left=0.25cm of b](c23);
     \vertex [above=0.28cm of c23](v23)[dot]{};
    \vertex [above=0.7cm of c](j1){$ 1$};
    \vertex [right=0.78cm of j1](j2){$2$};
    \vertex [right=0.52cm of j2](j3){$\sc 3$};
   	 \diagram*{(a) -- [thick] (b),(b) -- [thick] (j1),(v23) -- [thick] (j2),(b)--[thick](j3)};
    \end{feynman}  
  \end{tikzpicture}+\begin{tikzpicture}[baseline={([yshift=-0.8ex]current bounding box.center)}]\tikzstyle{every node}=[font=\small]    \begin{feynman}
    \vertex (a){$\sc 4$};
     \vertex [above=0.7cm of a](b)[dot]{};
     \vertex [left=0.7cm of b](c);
     \vertex [right=0.25cm of b](c23);
     \vertex [above=0.28cm of c23](v23)[dot]{};
    \vertex [above=0.7cm of c](j1){$1$};
    \vertex [right=0.58cm of j1](j2){$2$};
    \vertex [right=0.72cm of j2](j3){$\sc 3$};
   	 \diagram*{(a) -- [thick] (b),(b) -- [thick] (j1),(v23) -- [thick] (j2),(b)--[thick](j3)};
    \end{feynman}  
  \end{tikzpicture}={\npre([[ 1,2], \sc 3],\sc 4)\over p_{12}^2}+{\npre([ 1,[2,\sc 3]],\sc 4)\over p_{23}^2-m^2}\\
  &=\Big\langle{[[\ck 1,\ck 2],\ck 3]\over p_{12}^2}+{[\ck 1,[\ck 2,\ck 3]]\over p_{23}^2-m^2}\Big\rangle=\bigg( {\langle \ck 1 \star\ck 2 \star \ck 3+\ck 3 \star\ck 2 \star \ck 1 \rangle \over p_{12}^2}+{\langle \ck 1 \star\ck 2 \star\ck 3 +\ck 3 \star\ck 2 \star\ck 1\rangle \over p_{23}^2-m^2}\bigg)\, ,\nn
\end{align}
which agrees with eq.~\eqref{eq:a1234} after explicitly conducting the map. Here we use the nested commutator $\Gamma$ in $\npre$ to denote the BCJ numerator for the corresponding cubic graph, e.g. $\npre([[ 1,2], \sc 3],\sc 4)=\npre(1,2, \sc 3,\sc 4)-\npre(2,1, \sc 3,\sc 4)-\npre(\sc 3, 1,2,\sc 4)+\npre(\sc 3,2,1,\sc 4)$ denotes the BCJ numerator for first cubic graph in above equation.  

At five point, the pre-numerator is 
\begin{align}
   \npre(1,2, 3, \sc 4,\sc 5)=&\, \langle \ck 1\star \ck2\star \ck3\star \ck4\rangle \nn \\
   =&\, \bigg\langle T_{{(1)},{(2)},{(3)}}^{{(1234)}}+T_{{(2)},{(1)},{(3)}}^{{(1234)}}+T_{{(1)},{(3)},{(2)}}^{{(1234)}}+T_{{(3)},{(2)},{(1)}}^{{(1234)}}+T_{{(2)},{(3)},{(1)}}^{{(1234)}}+T_{{(3)},{(1)},{(2)}}^{{(1234)}} \nn \\
    &\quad -T_{{(1)},{(23)}}^{{(1234)}}-T_{{(23)},{(1)}}^{{(1234)}}-T_{{(12)},{(3)}}^{{(1234)}}-T_{{(3)},{(12)}}^{{(1234)}}-T_{{(13)},{(2)}}^{{(1234)}}-T_{{(2)},{(12)}}^{{(1234)}}+T_{{(123)}}^{{(1234)}}\bigg\rangle_\A \\
   =&\, \big\langle -T_{{(12)},{(3)}}^{{(1234)}} -T_{{(13)},{(2)}}^{{(1234)}}+T_{{(123)}}^{{(1234)}}\big\rangle_\A\nn\\
   =&\, {2p_4\Cdot F_{12}\Cdot p_4 p_{1}\Cdot F_3\Cdot p_4 \tr(t^{a_4}t^{a_5})\over (p_{14}^2-m^2)(p_{124}^2-m^2)}+{2p_4\Cdot F_{13}\Cdot p_4 p_{1}\Cdot F_2\Cdot p_{34} \tr(t^{a_4}t^{a_5})\over (p_{14}^2-m^2)(p_{134}^2-m^2)}-{p_4\Cdot F_{123}\Cdot p_4  \tr(t^{a_4}t^{a_5})\over p_{14}^2-m^2}\nn \, ,
   \end{align}
   and
   \begin{align}
   \npre(1,3, 2, \sc 4,\sc 5)=\npre(1,2, 3, \sc 4,\sc 5)|_{2\leftrightarrow 3}\, .
\end{align}
Then the amplitude is given by
\begin{align}
    \mathcal{A}(1,2,3,\sc 4,\sc{5})=2\, \mathsf{m}(12345|12345)\npre(1,2,3,\sc 4,\sc 5)+2\, \mathsf{m}(12345|13245)\npre(1,3,2,\sc 4,\sc 5)\, .
\end{align}
From the cubic graphs or nested commutators,  the amplitude is given as
\begin{align}
   &\mathcal{A}(1,2,3,\sc 4,\sc{5})=
    \begin{tikzpicture}[baseline={([yshift=-0.8ex]current bounding box.center)}]\tikzstyle{every node}=[font=\small]    
   \begin{feynman}
    \vertex (a){$\sc 5$};
     \vertex [above=0.7cm of a](b)[dot]{};
      \vertex [above=1.2cm of b](lb){$~~ 1 ~~~ 2~~~~3 ~~~~~\sc 4~$};
     \vertex [left=0.8cm of b](c);
     \vertex [right=0.25cm of b](c14);
     \vertex [right=0.5cm of b](c23);
     \vertex [above=0.46cm of c23](v23)[dot]{};
      \vertex [above=0.2cm of c14](v14)[dot]{};
    \vertex [above=0.9cm of c](j1);
    \vertex [right=0.6cm of j1](j4);
    \vertex [right=0.5cm of j4](j2);
    \vertex [right=0.5cm of j2](j3);
   	 \diagram*{(a) -- [thick] (b),(b) -- [thick] (j1),(v23) -- [thick] (j2),(v14) -- [thick] (j4),(b)--[thick](j3)};
    \end{feynman}  
  \end{tikzpicture}+ \begin{tikzpicture}[baseline={([yshift=-0.8ex]current bounding box.center)}]\tikzstyle{every node}=[font=\small]    
   \begin{feynman}
    \vertex (a){$\sc 5$};
     \vertex [above=0.7cm of a](b)[dot]{};
      \vertex [above=1.2cm of b](lb){$~ 1 ~~~ 2~~~~~3 ~~~~\sc 4~$};
     \vertex [left=0.8cm of b](c);
     \vertex [left=0.0cm of b](c14);
     \vertex [right=0.25cm of b](c23);
     \vertex [above=0.2cm of c23](v23)[dot]{};
      \vertex [above=0.46cm of c14](v14)[dot]{};
    \vertex [above=0.9cm of c](j1);
    \vertex [right=0.5cm of j1](j4);
    \vertex [right=0.6cm of j4](j2);
    \vertex [right=0.5cm of j2](j3);
   	 \diagram*{(a) -- [thick] (b),(b) -- [thick] (j1),(v14) -- [thick] (j2),(v14) -- [thick] (v23),(v14) -- [thick] (j4),(b)--[thick](j3)};
    \end{feynman}  
  \end{tikzpicture}+ \begin{tikzpicture}[baseline={([yshift=-0.8ex]current bounding box.center)}]\tikzstyle{every node}=[font=\small]    
   \begin{feynman}
    \vertex (a){$\sc 5$};
     \vertex [above=0.7cm of a](b)[dot]{};
      \vertex [above=1.2cm of b](lb){$~~1 ~~~ 2~~~~ 3 ~~~~~\sc 4~$};
     \vertex [left=0.8cm of b](c);
     \vertex [left=0.0cm of b](c14);
     \vertex [left=0.25cm of b](c23);
     \vertex [above=0.2cm of c23](v23)[dot]{};
      \vertex [above=0.46cm of c14](v14)[dot]{};
    \vertex [above=0.9cm of c](j1);
    \vertex [right=0.5cm of j1](j4);
    \vertex [right=0.6cm of j4](j2);
    \vertex [right=0.6cm of j2](j3);
   	 \diagram*{(a) -- [thick] (b),(b) -- [thick] (j1),(v23) -- [thick] (j2),(v14) -- [thick] (j4),(b)--[thick](j3)};
    \end{feynman}  
  \end{tikzpicture}+ \begin{tikzpicture}[baseline={([yshift=-0.8ex]current bounding box.center)}]\tikzstyle{every node}=[font=\small]    
   \begin{feynman}
    \vertex (a){$\sc 5$};
     \vertex [above=0.7cm of a](b)[dot]{};
      \vertex [above=1.2cm of b](lb){$~~1 ~~~~~ 2~~~~3 ~~~\sc 4~$};
     \vertex [left=0.8cm of b](c);
     \vertex [left=0.5cm of b](c14);
     \vertex [left=0.25cm of b](c23);
     \vertex [above=0.2cm of c23](v23)[dot]{};
      \vertex [above=0.46cm of c14](v14)[dot]{};
    \vertex [above=0.9cm of c](j1);
    \vertex [right=0.6cm of j1](j4);
    \vertex [right=0.5cm of j4](j2);
    \vertex [right=0.5cm of j2](j3);
   	 \diagram*{(a) -- [thick] (b),(b) -- [thick] (j1),(v23) -- [thick] (j2),(v14) -- [thick] (j4),(b)--[thick](j3)};
    \end{feynman}  
  \end{tikzpicture}+ \begin{tikzpicture}[baseline={([yshift=-0.8ex]current bounding box.center)}]\tikzstyle{every node}=[font=\small]    
   \begin{feynman}
    \vertex (a){$\sc 5$};
     \vertex [above=0.7cm of a](b)[dot]{};
      \vertex [above=1.2cm of b](lb){$~~1 ~~~~~ 2~~3 ~~~~~\sc 4~$};
     \vertex [left=0.8cm of b](c);
     \vertex [left=0.5cm of b](c14);
     \vertex [right=0.5cm of b](c23);
     \vertex [above=0.46cm of c23](v23)[dot]{};
      \vertex [above=0.46cm of c14](v14)[dot]{};
    \vertex [above=0.9cm of c](j1);
    \vertex [right=0.6cm of j1](j4);
    \vertex [right=0.5cm of j4](j2);
    \vertex [right=0.5cm of j2](j3);
   	 \diagram*{(a) -- [thick] (b),(b) -- [thick] (j1),(v23) -- [thick] (j2),(v14) -- [thick] (j4),(b)--[thick](j3)};
    \end{feynman}  
  \end{tikzpicture}
  \\
   &=\frac{\langle[\ck 1,[\ck 2,[\ck 3,\ck 4]]]\rangle}{(p_{34}^2{-}m^2) (p_{234}^2{-}m^2)}{+}\frac{\langle[\ck 1,[[\ck 2,\ck 3],\ck 4]]\rangle}{p_{23}^2 (p_{234}^2{-}m^2)}{+}\frac{\langle[[\ck 1,[\ck 2,\ck 3]],\ck 4]\rangle}{p_{23}^2 p_{123}^2}{+}\frac{\langle[[[\ck 1,\ck 2],\ck 3],\ck 4]\rangle}{p_{12}^2 p_{123}^2}{+}\frac{\langle[[\ck 1,\ck 2],[\ck 3,\ck 4]]\rangle}{p_{12}^2 (p_{34}^2{-}m^2)}\,\nn\\
&=\bigg({\langle (\Id_{\rm qs}+S)\ck 1\star \ck2\star \ck3\star \ck4\rangle \over (p_{34}^2{-}m^2) (p_{234}^2{-}m^2)}+\frac{\langle  (\Id_{\rm qs}+S)\ck 1\star \ck2\star \ck3\star \ck4\rangle - \langle  (\Id_{\rm qs}+S)\ck 1\star \ck3\star \ck2\star \ck4\rangle }{p_{23}^2 (p_{234}^2{-}m^2)}+\nn\\
& \ \ \frac{\langle  (\Id_{\rm qs}+S)(\ck 1\star \ck2\star \ck3\star \ck4 - \ck 1\star \ck3\star \ck2\star \ck4)\rangle }{p_{23}^2 p_{123}^2} +\frac{\langle  (\Id_{\rm qs}+S)\ck 1\star \ck2\star \ck3\star \ck4\rangle}{p_{12}^2 p_{123}^2}{+}\frac{\langle  (\Id_{\rm qs}+S)\ck 1\star \ck2\star \ck3\star \ck4\rangle}{p_{12}^2 (p_{34}^2{-}m^2)}\bigg)\,, \nn
\end{align}
where $S$ is the antipode, and we have used the fact that other fusion products all vanish under the evaluation map. 
The observation is  that there are only two independent BCJ numerators which agrees with the number of the minimal basis at five point. After collecting all the terms, we have 
\begin{align} \label{eq:A45}
   & \mathcal{A}(1,2,3,\sc 4,\sc{5})=2\Big({1\over (p_{34}^2{-}m^2) (p_{234}^2{-}m^2)}+\frac{1 }{p_{23}^2 (p_{234}^2{-}m^2)}+\frac{1 }{p_{23}^2 p_{123}^2}+\frac{1}{p_{12}^2 p_{123}^2}{+}\frac{1}{p_{12}^2 (p_{34}^2{-}m^2)}\Big)\nn \\
    &\times \Big({2p_4\Cdot F_{12}\Cdot p_4 p_{12}\Cdot F_3\Cdot p_4 \tr(t^{a_4}t^{a_5})\over (p_{14}^2-m^2)(p_{124}^2-m^2)}+{2p_4\Cdot F_{13}\Cdot p_4 p_{1}\Cdot F_2\Cdot p_{34} \tr(t^{a_4}t^{a_5})\over (p_{14}^2-m^2)(p_{134}^2-m^2)}-{p_4\Cdot F_{123}\Cdot p_4 \tr(t^{a_4}t^{a_5})\over (p_{14}^2-m^2)}\Big) \\
    &-\Big(\frac{p_{123}^2+ p_{234}^2{-}m^2}{p_{23}^2 p_{123}^2 (p_{234}^2{-}m^2)}\Big)\Big({2p_4\Cdot F_{13}\Cdot p_4 p_{13}\Cdot F_2\Cdot p_4 \tr(t^{a_4}t^{a_5})\over (p_{14}^2-m^2)(p_{134}^2-m^2)}+{2p_4\Cdot F_{12}\Cdot p_4 p_{1}\Cdot F_3\Cdot p_{24} \tr(t^{a_4}t^{a_5})\over (p_{14}^2-m^2)(p_{124}^2-m^2)}-{p_4\Cdot F_{132}\Cdot p_4  \tr(t^{a_4}t^{a_5})\over (p_{14}^2-m^2)}\Big)\, .\nn
\end{align}
It is illuminating to consider the following equivalent five-point amplitude, 
\begin{align}
    \mathcal{A}(\sc 4, 1,2,3,\sc 5)={\npre([\sc 4,[1,[2,3]]],\sc 5)\over p_{23}^2 p_{123}^2}{+}{\npre([\sc 4,[[1,2],3]],\sc 5)\over p_{12}^2 p_{123}^2}{+}{\npre([[\sc 4,[1,2]], 3],\sc 5)\over p_{12}^2 (p_{124}^2-m^2)}{+}{\npre([[[\sc 4,1],2],3],\sc 5)\over (p_{14}^2-m^2) (p_{124}^2)}{+}{\npre([[\sc 4,1],[2,3]],\sc 5)\over p_{23}^2 (p_{14}^2-m^2)}\, .
\end{align}
Following from eq.~\eqref{eq:000}, we have
\begin{align}
    \npre([[\sc 4,[1,2]], 3],\sc 5)=\npre([[[\sc 4,1],2],3],\sc 5)=\npre([[\sc 4,1],[2,3]],\sc 5)=0\, ,
\end{align}
which implies that
\begin{align}
    \mathcal{A}(\sc 4, 1,2,3,\sc 5)&={\npre([\sc 4,[1,[2,3]]],\sc 5)\over p_{23}^2 p_{123}^2}+{\npre([\sc 4,[[1,2],3]],\sc 5)\over p_{12}^2 p_{123}^2}\\
    &=\bigg(\hskip -3pt -{\langle (\Id_{\rm qs}+S)\ck 1\star \ck2\star \ck3\star \ck4\rangle -\langle (\Id_{\rm qs}+S)\ck 1\star \ck3\star \ck2\star \ck4\rangle \over p_{23}^2 p_{123}^2}-{\langle (\Id_{\rm qs}+S)\ck 1\star \ck2\star \ck3\star \ck4\rangle \over p_{12}^2 p_{123}^2}\bigg)\,.\nn
\end{align}

We will now consider amplitudes with three or more scalars, with the mapping rule given in eq.~\eqref{eq:Talpha} of the main text. 
The first example is a four-point amplitude ${\cal A}(\sc 1,3,\sc 2,\sc 4)$.  
The non-trivial pre-numerators are
\begin{align}
     \npre(\overline{1}, 3, \overline{2}, \overline{4}) &= \langle \ck 1 \star \ck 3 \star \ck 2 \rangle  = \langle T^{(132)}_{(3)}\rangle_{\A} \tr(t^{a_1} t^{a_2}t^{a_4}) \, ,\nn\\
     \npre(\overline{2}, 3,\overline{1}, \overline{4}) &= \langle \ck 2 \star \ck 3 \star \ck 1 \rangle  = \langle T^{(231)}_{(3)}\rangle_{\A} \tr(t^{a_2} t^{a_1}t^{a_4})\,,
\end{align}
where the evaluation map of the  generators can be read off directly from the musical diagrams
\begin{align}
\langle T^{(132)}_{(3)}\rangle_{\A}&=\bigg(\frac{2 p_{1}\Cdot F_{3}\Cdot p_{2}}{p_{12}^2-m^2}\bigg)
\qquad \quad
    \begin{tikzpicture}[baseline={([yshift=-0.8ex]current bounding box.center)}]\tikzstyle{every node}=[font=\small]    
   \begin{feynman}
    \vertex (l1)[]{}; 
    \vertex [below=0.4cm of l1](ls)[]{}; 
    \vertex [left=0.5cm of ls](rms)[]{$(\eta)$};
    \vertex [right=2.cm of ls](rs){}; 
    \vertex [left=0.5cm of l1](rm1)[]{$(\tau_1)$};
    \vertex [right=2.cm of l1](r1)[]{}; 
    \vertex [right=0.5cm of ls](vphif)[myblob2]{\white\small $\mathbf{1}$};
     \vertex [right=1.5cm of ls](v3)[myblob2]{\white\small ${\mathbf{2}}$};
      \vertex [right=1.0cm of l1](v5)[sb]{\white\textbf\small $\mathbf{3}$};
   	 \diagram*{(l1)--[thick](v5)--[thick](r1),(ls)--[thick](vphif)--[thick](v3)--[thick](rs)};
    \end{feynman}
    \end{tikzpicture} \, , 
\end{align}
and 
\begin{align}
\langle T^{(231)}_{(3)}\rangle_{\A}=\bigg(\frac{2 p_{2}\Cdot F_{3}\Cdot p_{1}}{p_{12}^2-m^2}\bigg)
\qquad \quad
    \begin{tikzpicture}[baseline={([yshift=-0.8ex]current bounding box.center)}]\tikzstyle{every node}=[font=\small]    
   \begin{feynman}
    \vertex (l1)[]{}; 
    \vertex [below=0.4cm of l1](ls)[]{}; 
    \vertex [left=0.5cm of ls](rms)[]{$(\eta)$};
    \vertex [right=2.cm of ls](rs){}; 
    \vertex [left=0.5cm of l1](rm1)[]{$(\tau_1)$};
    \vertex [right=2.cm of l1](r1)[]{}; 
    \vertex [right=0.5cm of ls](vphif)[myblob2]{\white\small $\mathbf{2}$};
     \vertex [right=1.5cm of ls](v3)[myblob2]{\white\small ${\mathbf{1}}$};
      \vertex [right=1.0cm of l1](v5)[sb]{\white\textbf\small $\mathbf{3}$};
   	 \diagram*{(l1)--[thick](v5)--[thick](r1),(ls)--[thick](vphif)--[thick](v3)--[thick](rs)};
    \end{feynman}
    \end{tikzpicture} \, .
    \end{align}
It is easy to see the antipode relation $\langle T^{(132)}_{(3)}\rangle_{\A}=-\langle T^{(231)}_{(3)}\rangle_{\A}$.
The evaluation map for all other  generators leads to vanishing pre-numerators 
\begin{align}
     \npre(\overline{1},  \overline{2},3, \overline{4})=\npre(\overline{2},  \overline{1},3, \overline{4})=\npre(3,\overline{1},  \overline{2}, \overline{4})=\npre(3,\overline{2},  \overline{1}, \overline{4})=0\, .
\end{align}
With these pre-numerators,  
the amplitude is given by
\begin{equation}
\begin{aligned}\label{eq:A4pm}
    {\cal A}(\sc 1,3,\sc 2,\sc 4)&=\mathsf{m}(1324|1324)\npre(\sc 1,3,\sc 2,\sc 4)+\mathsf{m}(1324|2314)\npre(\sc 2,3,\sc 1,\sc 4)\\
    &=\left({1\over p_{13}^2-m^2}+{1\over p_{23}^2-m^2}\right)\bigg(\frac{2 p_{1}\Cdot F_{3}\Cdot p_{2}}{p_{12}^2-m^2}\bigg)\tr([t^{a_1} ,t^{a_2}]t^{a_4})\,.
\end{aligned}
\end{equation}
In the nested commutator form, 
the same amplitude is given by summing over the relevant cubic graphs
\begin{align}\label{eq:A4ck2}
{\cal A}(\sc{1},3,\sc{2},\sc 4)&=
 \begin{tikzpicture}[baseline={([yshift=-0.8ex]current bounding box.center)}]\tikzstyle{every node}=[font=\small]    
   \begin{feynman}
    \vertex (a){$\sc 4$};
     \vertex [above=0.7cm of a](b)[dot]{};
     \vertex [left=0.7cm of b](c);
     \vertex [left=0.25cm of b](c23);
     \vertex [above=0.28cm of c23](v23)[dot]{};
    \vertex [above=0.7cm of c](j1){$\sc 1$};
    \vertex [right=0.78cm of j1](j2){$3$};
    \vertex [right=0.52cm of j2](j3){$\sc 2$};
   	 \diagram*{(a) -- [thick] (b),(b) -- [thick] (j1),(v23) -- [thick] (j2),(b)--[thick](j3)};
    \end{feynman}  
  \end{tikzpicture}+\begin{tikzpicture}[baseline={([yshift=-0.8ex]current bounding box.center)}]\tikzstyle{every node}=[font=\small]    
   \begin{feynman}
    \vertex (a){$\sc 4$};
     \vertex [above=0.7cm of a](b)[dot]{};
     \vertex [left=0.7cm of b](c);
     \vertex [right=0.25cm of b](c23);
     \vertex [above=0.28cm of c23](v23)[dot]{};
    \vertex [above=0.7cm of c](j1){$\sc 1$};
    \vertex [right=0.58cm of j1](j2){$3$};
    \vertex [right=0.72cm of j2](j3){$\sc 2$};
   	 \diagram*{(a) -- [thick] (b),(b) -- [thick] (j1),(v23) -- [thick] (j2),(b)--[thick](j3)};
    \end{feynman}  
  \end{tikzpicture}={\npre([[\sc 1,3],\sc 2],\sc 4)\over p_{13}^2-m^2}+{\npre([\sc 1,[3,\sc 2]],\sc 4)\over p_{23}^2-m^2}=\Big\langle{[[\ck 1,\ck 3],\ck 2]\over p_{13}^2-m^2}+{[\ck 1,[\ck 3,\ck 2]]\over p_{23}^2-m^2}\Big\rangle\nn\\
  &={\langle T^{(132)}_{(3)} \rangle_{\A} \tr(t^{a_1}t^{a_2}t^{a_4})+\langle T^{(231)}_{(3)}\rangle_{\A} \tr(t^{a_2}t^{a_1}t^{a_4}) \over p_{13}^2-m^2}+{\langle T^{(132)}_{(3)}\rangle_{\A} \tr(t^{a_1}t^{a_2}t^{a_4})+\langle T^{(231)}_{(3)}\rangle_{\A} \tr(t^{a_2}t^{a_1}t^{a_4}) \over p_{23}^2-m^2}\nn\\
  &={\langle T^{(132)}_{(3)}\rangle_{\A} \tr([t^{a_1},t^{a_2}]t^{a_4}) \over p_{13}^2-m^2}+{\langle T^{(132)}_{(3)}\rangle_{\A} \tr([t^{a_1},t^{a_2}]t^{a_4}) \over p_{23}^2-m^2}\, ,
\end{align}
where we have used the fact that $\langle T^{(123)}_{(3)}\rangle_{\A} =\langle T^{(312)}_{(3)}\rangle_{\A} =0$, and the antipode symmetry $\langle T^{(132)}_{(3)}\rangle_{\A}=-\langle T^{(231)}_{(3)}\rangle_{\A}$ is important to obtain the flavor structure constant (as a commutator) in the final expression. As expected, eq.~\eqref{eq:A4ck2} agrees with eq.~\eqref{eq:A4pm}. From this example, one can see the following interesting properties of the pre-numerators and BCJ numerators
 \begin{itemize}
 \item 
$\npre(\sc 1,3,\sc 2,\sc 4)|_{t^a\rightarrow \Id}=\npre([[\sc 1,3],\sc 2],\sc 4)|_{f^{abc}\rightarrow  1}$
 \item $\npre([[\sc 1,3],\sc 2],\sc 4)|_{f^{abc}\rightarrow  1}=\npre([\sc 1,[3,\sc 2]],\sc 4)|_{f^{abc}\rightarrow  1}$.
 \end{itemize}
 These are examples illustrating some general relations between pre-numerators and BCJ numerators discussed in the main text.

The second multi-scalar example is a five-point amplitude, $\mathcal{A}(\overline{1}, 4, \overline{2}, \overline{3},  \overline{5})$. The pre-numerators are computed from the kinematic algebra, for example,
\begin{equation}
\begin{aligned}
     \npre(\overline{1}, 4, \overline{2}, \overline{3}, \sc 5) &= \langle \ck 1 \star \ck 4 \star \ck 2 \star \ck 3\rangle  = \langle T^{(1423)}_{(4)}\rangle_{\A} \operatorname{tr}(t^{a_1} t^{a_2}t^{a_3}t^{a_5}) \,,
\end{aligned}
\end{equation}
where we have applied the mapping rule eq.~\eqref{eq:mapp} and 
\begin{equation}
   \langle T^{(1423)}_{(4)}\rangle_{\A}= \bigg(\frac{2 p_{1}\Cdot F_{4}\Cdot p_{23}}{p_{123}^2-m^2}\bigg) \qquad \quad \begin{tikzpicture}[baseline={([yshift=-0.8ex]current bounding box.center)}]\tikzstyle{every node}=[font=\small]    
   \begin{feynman}
    \vertex (l1)[]{}; 
    \vertex [below=0.4cm of l1](ls)[]{}; 
    \vertex [left=0.5cm of ls](rms)[]{$(\eta)$};
    \vertex [right=2.6cm of ls](rs){}; 
    \vertex [left=0.5cm of l1](rm1)[]{$(\tau_1)$};
    \vertex [right=2.6cm of l1](r1)[]{}; 
    \vertex [right=0.5cm of ls](vphif)[myblob2]{\white\small $\mathbf{1}$};
     \vertex [right=1.5cm of ls](v3)[myblob2]{\white\small ${\mathbf{2}}$};
      \vertex [right=2.1cm of ls](v4)[myblob2]{\white\small ${\mathbf{3}}$};
      \vertex [right=1.0cm of l1](v5)[sb]{\white\textbf\small $\mathbf{4}$};
   	 \diagram*{(l1)--[thick](v5)--[thick](r1),(ls)--[thick](vphif)--[thick](v3)--[thick](v4)--[thick](rs)};
    \end{feynman}
    \end{tikzpicture}    
    \begin{tikzpicture}[baseline={([yshift=-0.8ex]current bounding box.center)}]\tikzstyle{every node}=[font=\small]    
  \end{tikzpicture}\, .
\end{equation}
Analogous expressions are obtained for other numerators, {e.g.}
\begin{equation}
\begin{aligned}
     \npre(\overline{1}, \overline{2}, 4, \overline{3},\sc 5) =\langle T^{(1243)}_{(4)}\rangle \operatorname{tr}(t^{a_1}t^{a_2}t^{a_3}t^{a_5})=\bigg(\frac{2 p_{12}\Cdot F_{4} \Cdot p_{3}}{p_{{123}}^2-m^2}\bigg)\operatorname{tr}(t^{a_1}t^{a_2}t^{a_3}t^{a_5})\,.
\end{aligned}
\end{equation}
From these pre-numerators, the amplitude is given as
\begin{equation}
\begin{aligned} \label{eq:ex1}
     \mathcal{A}(\overline{1},  4, \overline{2}, \overline{3}, \overline{5})&=\sum_{\sigma\in S_4}\mathsf{m}(\sc{1}  4 \sc{2} \sc{3} \sc{5}|\sigma_{1}  \sigma_{2} \sigma_{3}\sigma_4 \sc{5})\npre(\sigma_{1},  \sigma_{2}, \sigma_{3},\sigma_4, \sc{5})\,  \\
     &=\sum_{\sigma\in S_3}\mathsf{m}(\sc{1}  4 \sc{2} \sc{3} \sc{5}|\sigma_{1}  \sigma_{2}4 \sigma_{3} \sc{5})\npre(\sigma_{1},  \sigma_{2},4, \sigma_{3}, \sc{5})+\sum_{\sigma\in S_3}\mathsf{m}(\sc{1}  4 \sc{2} \sc{3} \sc{5}|\sigma_{1} 4 \sigma_{2} \sigma_{3} \sc{5})\npre(\sigma_{1}, 4, \sigma_{2}, \sigma_{3}, \sc{5})\,
\end{aligned},
\end{equation}
where we use the fact that $\npre(\sigma_{1},  \sigma_{2}, \sigma_{3},4, \sc{5})=\npre(4,\sigma_{1},  \sigma_{2}, \sigma_{3}, \sc{5})=0$ and the propagator matrix $\mathsf{m}$ as the bi-adjoint scalar amplitudes is easy to obtain, for instance,
\begin{equation}
\begin{aligned}
     \mathsf{m}(\overline{1} 4 \overline{2} \overline{3}\overline{5}|\overline{1} \overline{2} 4 \overline{3}\overline{5})=-{1\over (p^2_{24}{-}m^2)(p^{2}_{124}{-}m^2)}-{1 \over (p^2_{24}{-}m^2)(p^{2}_{234}{-}m^2)} \, .
\end{aligned}
\end{equation}

The same amplitude can also be expressed in the nested commutator form
\begin{align}
   &\mathcal{A}(\sc{1},  4, \sc{2}, \sc{3}, \sc{5})=
    \begin{tikzpicture}[baseline={([yshift=-0.8ex]current bounding box.center)}]\tikzstyle{every node}=[font=\small]    
   \begin{feynman}
    \vertex (a){$\sc 5$};
     \vertex [above=0.7cm of a](b)[dot]{};
      \vertex [above=1.2cm of b](lb){$~~\sc 1 ~~~ 4~~~~\sc 2 ~~~~~\sc 3~$};
     \vertex [left=0.8cm of b](c);
     \vertex [right=0.25cm of b](c14);
     \vertex [right=0.5cm of b](c23);
     \vertex [above=0.46cm of c23](v23)[dot]{};
      \vertex [above=0.2cm of c14](v14)[dot]{};
    \vertex [above=0.9cm of c](j1);
    \vertex [right=0.6cm of j1](j4);
    \vertex [right=0.5cm of j4](j2);
    \vertex [right=0.5cm of j2](j3);
   	 \diagram*{(a) -- [thick] (b),(b) -- [thick] (j1),(v23) -- [thick] (j2),(v14) -- [thick] (j4),(b)--[thick](j3)};
    \end{feynman}  
  \end{tikzpicture}+ \begin{tikzpicture}[baseline={([yshift=-0.8ex]current bounding box.center)}]\tikzstyle{every node}=[font=\small]    
   \begin{feynman}
    \vertex (a){$\sc 5$};
     \vertex [above=0.7cm of a](b)[dot]{};
      \vertex [above=1.2cm of b](lb){$~\sc 1 ~~~ 4~~~~~\sc 2 ~~~~\sc 3~$};
     \vertex [left=0.8cm of b](c);
     \vertex [left=0.0cm of b](c14);
     \vertex [right=0.25cm of b](c23);
     \vertex [above=0.2cm of c23](v23)[dot]{};
      \vertex [above=0.46cm of c14](v14)[dot]{};
    \vertex [above=0.9cm of c](j1);
    \vertex [right=0.5cm of j1](j4);
    \vertex [right=0.6cm of j4](j2);
    \vertex [right=0.5cm of j2](j3);
   	 \diagram*{(a) -- [thick] (b),(b) -- [thick] (j1),(v14) -- [thick] (j2),(v14) -- [thick] (v23),(v14) -- [thick] (j4),(b)--[thick](j3)};
    \end{feynman}  
  \end{tikzpicture}+ \begin{tikzpicture}[baseline={([yshift=-0.8ex]current bounding box.center)}]\tikzstyle{every node}=[font=\small]    
   \begin{feynman}
    \vertex (a){$\sc 5$};
     \vertex [above=0.7cm of a](b)[dot]{};
      \vertex [above=1.2cm of b](lb){$~~\sc 1 ~~~ 4~~~~\sc 2 ~~~~~\sc 3~$};
     \vertex [left=0.8cm of b](c);
     \vertex [left=0.0cm of b](c14);
     \vertex [left=0.25cm of b](c23);
     \vertex [above=0.2cm of c23](v23)[dot]{};
      \vertex [above=0.46cm of c14](v14)[dot]{};
    \vertex [above=0.9cm of c](j1);
    \vertex [right=0.5cm of j1](j4);
    \vertex [right=0.6cm of j4](j2);
    \vertex [right=0.6cm of j2](j3);
   	 \diagram*{(a) -- [thick] (b),(b) -- [thick] (j1),(v23) -- [thick] (j2),(v14) -- [thick] (j4),(b)--[thick](j3)};
    \end{feynman}  
  \end{tikzpicture}+ \begin{tikzpicture}[baseline={([yshift=-0.8ex]current bounding box.center)}]\tikzstyle{every node}=[font=\small]    
   \begin{feynman}
    \vertex (a){$\sc 5$};
     \vertex [above=0.7cm of a](b)[dot]{};
      \vertex [above=1.2cm of b](lb){$~~\sc 1 ~~~~~ 4~~~~\sc 2 ~~~\sc 3~$};
     \vertex [left=0.8cm of b](c);
     \vertex [left=0.5cm of b](c14);
     \vertex [left=0.25cm of b](c23);
     \vertex [above=0.2cm of c23](v23)[dot]{};
      \vertex [above=0.46cm of c14](v14)[dot]{};
    \vertex [above=0.9cm of c](j1);
    \vertex [right=0.6cm of j1](j4);
    \vertex [right=0.5cm of j4](j2);
    \vertex [right=0.5cm of j2](j3);
   	 \diagram*{(a) -- [thick] (b),(b) -- [thick] (j1),(v23) -- [thick] (j2),(v14) -- [thick] (j4),(b)--[thick](j3)};
    \end{feynman}  
  \end{tikzpicture}+ \begin{tikzpicture}[baseline={([yshift=-0.8ex]current bounding box.center)}]\tikzstyle{every node}=[font=\small]    
   \begin{feynman}
    \vertex (a){$\sc 5$};
     \vertex [above=0.7cm of a](b)[dot]{};
      \vertex [above=1.2cm of b](lb){$~~\sc 1 ~~~~~ 4~~\sc 2 ~~~~~\sc 3~$};
     \vertex [left=0.8cm of b](c);
     \vertex [left=0.5cm of b](c14);
     \vertex [right=0.5cm of b](c23);
     \vertex [above=0.46cm of c23](v23)[dot]{};
      \vertex [above=0.46cm of c14](v14)[dot]{};
    \vertex [above=0.9cm of c](j1);
    \vertex [right=0.6cm of j1](j4);
    \vertex [right=0.5cm of j4](j2);
    \vertex [right=0.5cm of j2](j3);
   	 \diagram*{(a) -- [thick] (b),(b) -- [thick] (j1),(v23) -- [thick] (j2),(v14) -- [thick] (j4),(b)--[thick](j3)};
    \end{feynman}  
  \end{tikzpicture}
  \\
   &=\frac{\langle[\ck 1,[\ck 4,[\ck 2,\ck 3]]]\rangle}{(p_{23}^2{-}m^2) (p_{234}^2{-}m^2)}{+}\frac{\langle[\ck 1,[[\ck 4,\ck 2],\ck 3]]\rangle}{(p_{24}^2{-}m^2) (p_{234}^2{-}m^2)}{+}\frac{\langle[[\ck 1,[\ck 4,\ck 2]],\ck 3]\rangle}{(p_{24}^2{-}m^2) (p_{124}^2{-}m^2)}{+}\frac{\langle[[[\ck 1,\ck 4],\ck 2],\ck 3]\rangle}{(p_{14}^2{-}m^2) (p_{124}^2{-}m^2)}{+}
\frac{\langle[[\ck 1,\ck 4],[\ck 2,\ck 3]]\rangle}{(p_{14}^2{-}m^2) (p_{23}^2{-}m^2)}\,.\nn
\end{align}
After writing out explicitly each term, we obtain precisely eq.~\eqref{eq:ex1}, which shows the equivalence between these two representations. 
Each BCJ numerator can be evaluated directly from the algebraic fusion product and the evaluation map, {e.g.}
\begin{align}
   \npre([\sc 1, [4,[\sc 2,\sc 3]]],\sc 5)&= \langle[\ck 1,[\ck 4,[\ck 2,\ck 3]]]\rangle\nn\\
   &=\langle\ck 1\star\ck 4\star\ck 2\star\ck 3\rangle-\langle\ck 1\star\ck 4\star\ck 3\star\ck 2\rangle-\langle\ck 1\star\ck 2\star\ck 3\star\ck 4\rangle+\langle\ck 1\star\ck 3\star\ck 2\star\ck 4\rangle\nn\\
   &\quad -(\langle\ck 4\star\ck 2\star\ck 3\star \ck 1\rangle-\langle\ck 4\star\ck 3\star\ck 2\star \ck 1\rangle-\langle\ck 2\star\ck 3\star\ck 4\star \ck 1\rangle+\langle\ck 3\star\ck 2\star\ck 4\star \ck 1\rangle)\nn\\
   &=\langle T^{(1423)}_{(4)}\rangle_\A \tr(t^{a_1}[t^{a_2},t^{a_3}]t^{a_5})-\langle T^{(3241)}_{(4)}\rangle_\A \tr([t^{a_3},t^{a_2}]t^{a_1}t^{a_5})\nn\\
   &=\langle T^{(1423)}_{(4)}\rangle_\A \tr([t^{a_1},[t^{a_2},t^{a_3}]]t^{a_5})\, ,
\end{align}
where we have used the mapping rule and its properties in the main text to obtain the final result. Once again the pre-numerators and BCJ numerators obey many interesting properties, e.g.
 \begin{itemize}
 \item 
$\npre(\sc 1,4,\sc 2,\sc 3,\sc 5)|_{t^a\rightarrow \Id}=\npre([\sc 1,4,\sc 2,\sc 3],\sc 5)|_{f^{abc}\rightarrow  1}$.
 \item $\npre([[\sc 1,4],[\sc 2,\sc 3]],\sc 5)|_{f^{abc}\rightarrow 1}=\npre([\sc 1,[4.[\sc 2,\sc 3]]],\sc 5)|_{f^{abc}\rightarrow 1}$.
 \end{itemize}
The amplitude can then be expressed as 
 \begin{align}
     \mathcal{A}(\sc{1},  4, \sc{2}, \sc{3}, \sc{5})&=\frac{\langle T^{(1423)}_{(4)}\rangle_{\A} \tr([t^{a_1},[t^{a_2},t^{a_3}]]t^{a_5})}{(p_{23}^2{-}m^2) (p_{234}^2{-}m^2)}{+}\frac{\langle T^{(1342)}_{(4)}\rangle \tr([t^{a_1},[t^{a_2},t^{a_3}]]t^{a_5})}{(p_{24}^2{-}m^2) (p_{234}^2{-}m^2)}{+}
\frac{\langle T^{(1423)}_{(4)}\rangle \tr([t^{a_1},[t^{a_2},t^{a_3}]]t^{a_5})}{(p_{14}^2{-}m^2) (p_{23}^2{-}m^2)}\nn\\
     &\quad \ {+}\frac{\langle T^{(1342)}_{(4)}\rangle \tr([[t^{a_1},t^{a_2}],t^{a_3}]t^{a_5})}{(p_{24}^2{-}m^2) (p_{124}^2{-}m^2)}{+}\frac{\langle T^{(1423)}_{(4)}\rangle \tr([[t^{a_1},t^{a_2}],t^{a_3}]t^{a_5})}{(p_{14}^2{-}m^2) (p_{124}^2{-}m^2)}\, .
 \end{align}

Similar expressions are found for the form factor $\mathcal{F}(\overline{1}, 4, \overline{2}, \overline{3})$, 
\begin{equation}\label{eq:F4prenum}
\begin{aligned}
    \mathcal{F}(\overline{1},4, \overline{2}, \overline{3})=&\,\mathsf{m}_{\cal F}(\overline{1}  4 \overline{2} \overline{3}|\overline{1}  4 \overline{2} \overline{3})\npre_{\cal F}(\overline{1},  4, \overline{2}, \overline{3})+\mathsf{m}_{\cal F}(\overline{1}  4 \overline{2} \overline{3}|\overline{1}  4 \overline{3} \overline{2})\npre_{\cal F}(\overline{1},  4, \overline{3}, \overline{2})\\
    &+ \, \mathsf{m}_{\cal F}(\overline{1}  4 \overline{2} \overline{3}|\overline{1} \overline{3} 4 \overline{2})\npre_{\cal F}(\overline{1}, \overline{3}, 4, \overline{2})+\mathsf{m}_{\cal F}(\overline{1}  4 \overline{2} \overline{3}|\overline{1} \overline{2} 4 \overline{3})\npre_{\cal F}(\overline{1}, \overline{2}, 4, \overline{3})\, ,
\end{aligned}
\end{equation}
where the pre-numerators are obtained from those of the amplitude in eq.~\eqref{eq:ex1} by replacing $m^2 \rightarrow q^2$. However, the propagator matrix $\mathsf{m}_{\cal F}$ contains more terms, {e.g.}
\begin{equation}
\begin{aligned}
     \mathsf{m}_{\cal F}(\overline{1} 4 \overline{2} \overline{3}|\overline{1} \overline{2} 4 \overline{3})=&-{1\over (p^2_{24}{-}m^2)(p^{2}_{124}{-}m^2)}-{1 \over (p^2_{24}{-}m^2)(p^{2}_{234}{-}m^2)}-{1\over (p^2_{13}{-}m^2)(p^{2}_{24}{-}m^2)}-{1\over (p^2_{13}{-}m^2)(p^{2}_{134}{-}m^2)} .
\end{aligned}
\end{equation}
The nested commutator form for $  \mathcal{F}(\overline{1},4, \overline{2}, \overline{3})$ is  obtained from the following cubic graphs
\begin{align}
    {\cal F}(\sc{1},4, \sc{2}, \sc{3})&= ~~~~\begin{tikzpicture}[baseline={([yshift=-0.8ex]current bounding box.center)}]\tikzstyle{every node}=[font=\small]    
   \begin{feynman}
    \vertex (a){$q$};
     \vertex [above=0.7cm of a](b)[fp]{};
      \vertex [above=1.2cm of b](lb){$~~\sigma_1 ~~ \sigma_4~~ \sigma_2 ~~\sigma_3~$};
     \vertex [left=0.8cm of b](c);
     \vertex [right=0.25cm of b](c14);
     \vertex [right=0.5cm of b](c23);
     \vertex [above=0.46cm of c23](v23)[dot]{};
      \vertex [above=0.2cm of c14](v14)[dot]{};
    \vertex [above=0.9cm of c](j1);
    \vertex [right=0.6cm of j1](j4);
    \vertex [right=0.5cm of j4](j2);
    \vertex [right=0.5cm of j2](j3);
   	 \diagram*{(a) -- [blue, ultra thick] (b),(b) -- [thick] (j1),(v23) -- [thick] (j2),(v14) -- [thick] (j4),(b)--[thick](j3)};
    \end{feynman}  
  \end{tikzpicture}~~~~~~~+~~~~ \begin{tikzpicture}[baseline={([yshift=-0.8ex]current bounding box.center)}]\tikzstyle{every node}=[font=\small]    
   \begin{feynman}
    \vertex (a){$q$};
     \vertex [above=0.7cm of a](b)[fp]{};
      \vertex [above=1.2cm of b](lb){$~\sigma_1 ~~~ \sigma_4~~~\sigma_2 ~~~~\sigma_3~$};
     \vertex [left=0.8cm of b](c);
     \vertex [left=0.0cm of b](c14);
     \vertex [right=0.25cm of b](c23);
     \vertex [above=0.2cm of c23](v23)[dot]{};
      \vertex [above=0.46cm of c14](v14)[dot]{};
    \vertex [above=0.9cm of c](j1);
    \vertex [right=0.5cm of j1](j4);
    \vertex [right=0.6cm of j4](j2);
    \vertex [right=0.5cm of j2](j3);
   	 \diagram*{(a) -- [blue, ultra thick] (b),(b) -- [thick] (j1),(v14) -- [thick] (j2),(v14) -- [thick] (v23),(v14) -- [thick] (j4),(b)--[thick](j3)};
    \end{feynman}  
  \end{tikzpicture}~~~~~~+~~~~
  \begin{tikzpicture}[baseline={([yshift=-0.8ex]current bounding box.center)}]\tikzstyle{every node}=[font=\small]    
   \begin{feynman}
    \vertex (a){$q$};
     \vertex [above=0.7cm of a](b)[fp]{};
      \vertex [above=1.2cm of b](lb){$~~\sigma_1 ~~~~ \sigma_4~~\sigma_2 ~~\sigma_3~$};
     \vertex [left=0.8cm of b](c);
     \vertex [left=0.5cm of b](c14);
     \vertex [right=0.5cm of b](c23);
     \vertex [above=0.46cm of c23](v23)[dot]{};
      \vertex [above=0.46cm of c14](v14)[dot]{};
    \vertex [above=0.9cm of c](j1);
    \vertex [right=0.6cm of j1](j4);
    \vertex [right=0.5cm of j4](j2);
    \vertex [right=0.5cm of j2](j3);
   	 \diagram*{(a) -- [blue, ultra thick] (b),(b) -- [thick] (j1),(v23) -- [thick] (j2),(v14) -- [thick] (j4),(b)--[thick](j3)};
    \end{feynman}  
  \end{tikzpicture}\\
  &=\sum_{\sigma\in {\rm cyc}(1423)}\bigg(\frac{\langle \ck {\sigma_1}\star [\ck {\sigma_4},[\ck {\sigma_2},\ck {\sigma_3}]]]\rangle}{(p_{\sigma_2\sigma_3}^2{-}m^2) (p_{\sigma_2\sigma_3\sigma_4}^2{-}m^2)}+\frac{\langle \ck {\sigma_1}\star [[\ck {\sigma_4},\ck {\sigma_2}],\ck {\sigma_3}]\rangle}{(p_{\sigma_2\sigma_4}^2{-}m^2) (p_{\sigma_2\sigma_3\sigma_4}^2{-}m^2)}+\frac{\langle [[\ck {\sigma_1}, \ck {\sigma_4}],[\ck {\sigma_2},\ck {\sigma_3}]]\rangle}{2(p_{\sigma_1\sigma_4}^2{-}m^2) (p_{\sigma_2\sigma_3}^2{-}m^2)}\bigg)\, ,\nn
\end{align}
which equals to eq.~\eqref{eq:F4prenum} after expanding the commutators. Similarly as the amplitude, each cubic graph is evaluated by the fusion product and evaluation map, e.g.
\begin{align}
    \npre_{\cal F}(\sc 1,[4,[\sc 2,\sc 3]])&=\langle \ck {1}\star [\ck {4},[\ck {2},\ck {3}]]]\rangle\\
    &=\langle\ck 1\star\ck 4\star\ck 2\star\ck 3\rangle-\langle\ck 1\star\ck 4\star\ck 3\star\ck 2\rangle-\langle\ck 1\star\ck 2\star\ck 3\star\ck 4\rangle+\langle\ck 1\star\ck 3\star\ck 2\star\ck 4\rangle\nn\\
    &=\langle T^{(1423)}_{(4)}\rangle_\FF \tr(t^{a_1}[t^{a_2},t^{a_3}])\, ,
\end{align}
and the numerators obey the condition  \eqref{eq:ff-relation}, for example, 
\begin{align}
    \npre_{\cal F}([\sc 1,4],[\sc 2,\sc 3])=\npre_{\cal F}(\sc 1,[4,[\sc 2,\sc 3]]) \, .
\end{align}
Then the form factor is given by
\begin{align}
    {\cal F}(\sc 1,4,\sc 2,\sc 3)=&\, {\langle T^{(1423)}_{(4)}\rangle_\FF \tr(t^{a_1}[t^{a_2},t^{a_3}])\over (p_{23}^2{-}m^2) (p_{234}^2{-}m^2)}+{\langle T^{(1342)}_{(4)}\rangle_\FF \tr(t^{a_3}[t^{a_1},t^{a_2}])\over (p_{24}^2{-}m^2) (p_{124}^2{-}m^2)}+{\langle T^{(1423)}_{(4)}\rangle_\FF \tr(t^{a_2}[t^{a_3},t^{a_1}])\over (p_{14}^2{-}m^2) (p_{134}^2{-}m^2)}\nn\\
    &+{\langle T^{(1342)}_{(4)}\rangle_\FF \tr(t^{a_1}[t^{a_2},t^{a_3}])\over (p_{24}^2{-}m^2) (p_{234}^2{-}m^2)}+{\langle T^{(1423)}_{(4)}\rangle_\FF \tr(t^{a_3}[t^{a_1},t^{a_2}])\over (p_{14}^2{-}m^2) (p_{124}^2{-}m^2)}+{\langle T^{(1342)}_{(4)}\rangle_\FF \tr(t^{a_2}[t^{a_3},t^{a_1}])\over (p_{13}^2{-}m^2) (p_{134}^2{-}m^2)}\nn\\
    &+{\langle T^{(1423)}_{(4)}\rangle_\FF \tr(t^{a_1}[t^{a_2},t^{a_3}])\over (p_{23}^2{-}m^2) (p_{14}^2{-}m^2)}+{\langle T^{(1342)}_{(4)}\rangle_\FF \tr(t^{a_2}[t^{a_3},t^{a_1}])\over (p_{23}^2{-}m^2) (p_{14}^2{-}m^2)}.
\end{align}

The other five point amplitude $\mathcal{A}(\overline{1},3,4,\overline{2},\overline{5})$ (and form factor $\mathcal{F}(\overline{1},3,4,\overline{2})$) can be obtained similarly either from pre-numerators or the cubic graphs. Just to give an example, 
\begin{align}\label{eq:3s2gnpre}
       \npre(\sc 1, 3, 4, \sc 2,\sc 5)= &\langle \ck 1\star\ck 3\star\ck 4\star \ck 2 \rangle =\langle -T^{(1342)}_{(34)}+T^{(1342)}_{(3),(4)}+T^{(1 3 4 2)}_{(4),(3)}\rangle_{\A} \, {\rm tr}(t^{a_1}t^{a_2} t^{a_5})\,,
\end{align}
and other BCJ numerators are either vanishing or related to this one by relabeling. 
The evaluation maps for algebraic generators in eq.~\eqref{eq:3s2gnpre} are obtained from the musical diagrams
\begin{align}
    \begin{tikzpicture}[baseline={([yshift=-0.8ex]current bounding box.center)}]\tikzstyle{every node}=[font=\small]    
   \begin{feynman}
    \vertex (l1)[]{}; 
    \vertex [below=0.4cm of l1](ls)[]{}; 
    \vertex [left=0.5cm of ls](rms)[]{$(\eta)$};
    \vertex [right=2.6cm of ls](rs){}; 
    \vertex [left=0.5cm of l1](rm1)[]{$(\tau_1)$};
    \vertex [right=2.6cm of l1](r1)[]{}; 
    \vertex [right=0.5cm of ls](vphif)[myblob2]{\white\small $\mathbf{1}$};
      \vertex [right=2.1cm of ls](v4)[myblob2]{\white\small ${\mathbf{2}}$};
      \vertex [right=1.0cm of l1](v5)[sb]{\white\textbf\small $\mathbf{3}$};
      \vertex [right=1.6cm of l1](v3)[sb]{\white\small ${\mathbf{4}}$};
   	 \diagram*{(l1)--[thick](v5)--[thick](v3)--[thick](r1),(ls)--[thick](vphif)--[thick](v4)--[thick](rs)};
    \end{feynman}
    \end{tikzpicture} && \begin{tikzpicture}[baseline={([yshift=-0.8ex]current bounding box.center)}]\tikzstyle{every node}=[font=\small]    
   \begin{feynman}
    \vertex (l1)[]{}; 
    \vertex [below=0.4cm of l1](ls)[]{}; \vertex [above=0.3cm of l1](l2)[]{};
    \vertex [left=0.5cm of ls](rms)[]{$(\eta)$};
    \vertex [right=2.6cm of ls](rs){}; 
    \vertex [left=0.5cm of l1](rm1)[]{$(\tau_1)$};
    \vertex [right=2.6cm of l1](r1)[]{};
     \vertex [left=0.5cm of l2](rm2)[]{$(\tau_2)$};
    \vertex [right=2.6cm of l2](r2)[]{};
    \vertex [right=0.5cm of ls](vphif)[myblob2]{\white\small $\mathbf{1}$};
      \vertex [right=2.1cm of ls](v4)[myblob2]{\white\small ${\mathbf{2}}$};
      \vertex [right=1.0cm of l1](v5)[sb]{\white\textbf\small $\mathbf{3}$};
      \vertex [right=1.6cm of l2](v3)[sb]{\white\small ${\mathbf{4}}$};
   	 \diagram*{(l2)--[thick](v3)--[thick](r2),(l1)--[thick](v5)--[thick](r1),(ls)--[thick](vphif)--[thick](v4)--[thick](rs)};
    \end{feynman}
    \end{tikzpicture} && 
    \begin{tikzpicture}[baseline={([yshift=-0.8ex]current bounding box.center)}]\tikzstyle{every node}=[font=\small]    
   \begin{feynman}
    \vertex (l1)[]{}; 
    \vertex [below=0.4cm of l1](ls)[]{}; \vertex [above=0.3cm of l1](l2)[]{};
    \vertex [left=0.5cm of ls](rms)[]{$(\eta)$};
    \vertex [right=2.6cm of ls](rs){}; 
    \vertex [left=0.5cm of l1](rm1)[]{$(\tau_1)$};
    \vertex [right=2.6cm of l1](r1)[]{};
     \vertex [left=0.5cm of l2](rm2)[]{$(\tau_2)$};
    \vertex [right=2.6cm of l2](r2)[]{};
    \vertex [right=0.5cm of ls](vphif)[myblob2]{\white\small $\mathbf{1}$};
      \vertex [right=2.1cm of ls](v4)[myblob2]{\white\small ${\mathbf{2}}$};
      \vertex [right=1.0cm of l2](v5)[sb]{\white\textbf\small $\mathbf{3}$};
      \vertex [right=1.6cm of l1](v3)[sb]{\white\small ${\mathbf{4}}$};
   	 \diagram*{(l2)--[thick](v5)--[thick](r2),(l1)--[thick](v3)--[thick](r1),(ls)--[thick](vphif)--[thick](v4)--[thick](rs)};
    \end{feynman}
    \end{tikzpicture} \nn\\
    \langle T^{(1342)}_{(34)}\rangle=\frac{2p_{{1}}\Cdot F_{34} \Cdot p_{{2}}}{p_{{12}}^2-m^2}&& \langle T^{(1342)}_{(3),(4)}\rangle =\frac{4p_{13}\Cdot F_{3}\Cdot p_{2}\, p_{13} \Cdot F_{4}\Cdot p_{{2}}}{(p_{{12}}^2-m^2)(p_{123}^2-m^2)}&&\langle T^{(1342)}_{(4),(3)}\rangle=\frac{4 p_{{1}}\Cdot F_{4}\Cdot p_{{2}}\,p_{14}\Cdot F_{3} \Cdot p_{{2}}}{(p_{{12}}^2-m^2)(p_{124}^2-m^2)}
\end{align}
So that the amplitude is given by
\begin{equation}
\begin{aligned}
     {\cal A}(\sc 1, 3, 4, \sc 2,\sc 5)=&\sum_{\rho\in S_2}\mathsf{m}(\sc{1} 3 4 \sc{2}  \sc{5}|\sc{1} \rho_3 \rho_4 \sc{2} \sc{5})\npre(\sc{1}, \rho_3, \rho_4, \sc{2}, \sc{5})+\sum_{\rho\in S_2}\mathsf{m}(\sc{1} 3 4 \sc{2}  \sc{5}|\sc{2} \rho_3 \rho_4 \sc{1} \sc{5})\npre(\sc{2}, \rho_3, \rho_4, \sc{1}, \sc{5})\\
     =&\,{\rm tr}(t^{a_1}t^{a_2} t^{a_5})\sum_{\rho\in S_2}\mathsf{m}(\sc{1} 3 4 \sc{2}  \sc{5}|\sc{1} \rho_3 \rho_4 \sc{2} \sc{5})\langle -T^{(1 \rho_3 \rho_4 2)}_{(\rho_3 \rho_4)}+T^{(1 \rho_3 \rho_4 2)}_{(\rho_3),(\rho_4)}+T^{(1 \rho_3 \rho_4 2)}_{(\rho_4),(\rho_3)}\rangle_{\A} \\
     &{\rm tr}(t^{a_2}t^{a_1} t^{a_5})\sum_{\rho\in S_2}\mathsf{m}(\sc{1} 3 4 \sc{2}  \sc{5}|\sc{2} \rho_3 \rho_4 \sc{1} \sc{5})\langle -T^{(2 \rho_3 \rho_4 1)}_{(\rho_3 \rho_4)}+T^{(2 \rho_3 \rho_4 1)}_{(\rho_3),(\rho_4)}+T^{(2 \rho_3 \rho_4 1)}_{(\rho_4),(\rho_3)}\rangle_{\A}\\
     =&\,{\rm tr}([t^{a_1},t^{a_2}] t^{a_5})\sum_{\rho\in S_2}\mathsf{m}(\sc{1} 3 4 \sc{2}  \sc{5}|\sc{1} \rho_3 \rho_4 \sc{2} \sc{5})\langle -T^{(1 \rho_3 \rho_4 2)}_{(\rho_3 \rho_4)}+T^{(1 \rho_3 \rho_4 2)}_{(\rho_3),(\rho_4)}+T^{(1 \rho_3 \rho_4 2)}_{(\rho_4),(\rho_3)}\rangle_{\A}\, ,
\end{aligned}
\end{equation}
where we have used the antipode symmetry in eq.~\eqref{eq:reverseSym} in the main text and the reflection property of the propagator matrix, {e.g.} $\mathsf{m}(13425|13425)={-}\mathsf{m}(13425|24315)$.

\section{D. More on colour-kinematic duality in form factor}

The universality of the kinematic Hopf algebra is further established for the form factors of more general operators. In the section, we consider operators that are built out of $h$ bi-adjoint scalars. Explicitly, they are defined as 
\begin{equation}
\begin{aligned}
{\rm Tr}{(\phi^h)}=\tr(t_c^{A_1}\cdots t_c^{A_h})\tr(t^{a_1}\cdots t^{a_h})\phi^{a_1,A_1}\cdots\phi^{a_h,A_h}\, .
\end{aligned}
\end{equation}
We find that the kinematic Hopf algebra we proposed in the main text for the the form factors with operator ${\rm Tr}{(\phi^2)}$ (i.e. $h=2$) can apply directly for these more general form factors, which lead to a novel closed-form expression for the ${\rm Tr}{(\phi^h)}$ form factors,
\begin{align}
 \mathcal{F}_{{\rm Tr}{(\phi^h)}}(\sigma)=&\sum_{\Gamma\in \graphset_{\sigma}^{(h)}} \qquad 
 {\langle\widehat\npre(\Gamma_{1})\star\cdots\star \widehat\npre(\Gamma_{h})\rangle \over d_{\Gamma_{1}}\cdots d_{\Gamma_{h}}} \begin{tikzpicture}[baseline={([yshift=-0.8ex]current bounding box.center)}]\tikzstyle{every node}=[font=\small]    
   \begin{feynman}
    \vertex (a)[]{$q$};
     \vertex [above=0.6cm of a](b)[fp]{};
     \vertex [left=1.1cm of b](c);
     \vertex [left=0.62cm of b](cL);
     \vertex [above=0.33cm of cL](vL)[HV]{\tiny$\Gamma_{1}$};
     \vertex [right=1.23cm of vL](vR)[HV]{\tiny$\Gamma_{h}$};
      \vertex [right=0.62cm of vL](vm){$\cdots$};
     \vertex [above=0.6cm of vL](t1){$\sigma_{1}$};
     \vertex [above=0.6cm of vR](t2){$~~\sigma_{h}$};
    \vertex [above=1.1cm of c](j1){};
    \vertex [right=.8cm of j1](j2){};
     \vertex [right=.55cm of j2](j3){};
    \vertex [right=0.8cm of j3](j4){};
   	 \diagram*{(a) -- [blue, ultra thick] (b),(b)--[thick](vL) -- [thick] (j1),(vL) -- [thick] (j2),(vR)--[thick](j3),(b)--[thick](vR)--[thick](j4)};
    \end{feynman}  
  \end{tikzpicture} \, , 
\end{align}
where the $\graphset_{\sigma}^{(h)}$ denotes all the graphs with colour ordering $\sigma$ that have $h$ components of cubic graphs $\Gamma_{i}, i\in [1,h]$. Although the vertex connecting with the operator is beyond the cubic vertices, it is just a fusion product of multi-algebraic generators with a given order. All the fusion products and propagators are identical to those in the form factor $\mathcal{F}_{{\rm Tr}{(\phi^2)}}$.

Furthermore, the numerators obey a variety of interesting relations similar to those discussed in the main text. In particular, we would like to mention the relation that are beyond the Jacobi identity 
\begin{align} \label{eq:ff-relation2}
&\sum_{i=1}^{h}  \langle\widehat\npre(\Gamma_1)\star \cdots \star \npre([\Gamma_i,j])\star \cdots \star \widehat\npre(\Gamma_h)\rangle =\sum_{i=1}^h \begin{tikzpicture}[baseline={([yshift=-0.8ex]current bounding box.center)}]\tikzstyle{every node}=[font=\small]    
   \begin{feynman}
    \vertex (a)[]{$q$};
     \vertex [above=0.6cm of a](b)[fp]{};
     \vertex [above=0.4cm of b](b1)[dot]{};
     \vertex [left=1.7cm of b](c);
     \vertex [left=1.22cm of b](cL);
     \vertex [above=0.63cm of cL](vL)[HV]{\tiny$\Gamma_{1}$};
     \vertex [right=2.23cm of vL](vR)[HV]{\tiny$\Gamma_{h}$};
      \vertex [right=1.02cm of vL](vm)[HV]{\tiny$\Gamma_{i}$};
       \vertex [right=0.5cm of vL](vdot1){$\cdots$};
        \vertex [right=0.5cm of vm](vdot2){$~~~\cdots$};
     \vertex [above=0.6cm of vL](t1){$\sigma_{1}~$};
     \vertex [above=0.6cm of vR](t2){$~~\sigma_{h}$};
     \vertex [above=0.6cm of vm](t3){$\sigma_{i}$};
    \vertex [above=1.4cm of c](j1){};
    \vertex [right=.7cm of j1](j2){};
     \vertex [right=.55cm of j2](j3){};
    \vertex [right=0.5cm of j3](j4){};
     \vertex [right=0.4cm of j4](j5){$j$};
      \vertex [right=0.4cm of j5](j6){};
       \vertex [right=0.6cm of j6](j7){};
   	 \diagram*{(a) -- [blue, ultra thick] (b),(b)--[thick](vL) -- [thick] (j1),(vL) -- [thick] (j2),(b1)--[thick](vm)--[thick](j3),(vm)--[thick](j4),(b)--[thick](b1)--[thick](j5), (vR)--[thick](j6),(b)--[thick](vR)--[thick](j7)};
    \end{feynman}  
  \end{tikzpicture}=0\, ,
\end{align}
where the $j$ line can be either a gluon or a scalar. This relation has been verified explicitly up to six points for the form factors with operators ${\rm Tr}(\phi^h)$ up to $h=6$. The kinematic Hopf algebra for these more general operators is of great significance for exploring the colour-kinematic duality for theories with higher-dimensional operators.  Below we will give a few concrete examples to illustrate our construction. 

Let us begin with the first non-trivial case: the four-point form factor of ${\rm Tr}(\phi^3)$
\begin{align}
 {\cal F}(\sc{1},\sc{2}, \sc{3},4)&= ~\begin{tikzpicture}[baseline={([yshift=-0.8ex]current bounding box.center)}]\tikzstyle{every node}=[font=\small]    
   \begin{feynman}
    \vertex (a){$q$};
     \vertex [above=0.7cm of a](b)[fp]{};
      \vertex [above=1.2cm of b](lb){$\sc 1 ~~~~~~\sc 2~~~\sc 3 ~~~~4$};
     \vertex [left=0.8cm of b](c);
     \vertex [right=0.25cm of b](c14);
     \vertex [right=0.5cm of b](c23);
     \vertex [above=0.46cm of c23](v23)[dot]{};
      \vertex [above=0.2cm of c14](v14)[]{};
    \vertex [above=0.9cm of c](j1);
    \vertex [right=0.8cm of j1](j4);
    \vertex [right=0.3cm of j4](j2);
    \vertex [right=0.5cm of j2](j3);
   	 \diagram*{(a) -- [blue, ultra thick] (b),(b) -- [thick] (j1),(v23) -- [thick] (j2),(b) -- [thick] (j4),(b)--[thick](j3)};
    \end{feynman}  
  \end{tikzpicture} + \begin{tikzpicture}[baseline={([yshift=-0.8ex]current bounding box.center)}]\tikzstyle{every node}=[font=\small]    
   \begin{feynman}
    \vertex (a){$q$};
     \vertex [above=0.7cm of a](b)[fp]{};
      \vertex [above=1.2cm of b](lb){$\sc 2 ~~~~~~\sc 3~~~ 4 ~~~~\sc 1$};
     \vertex [left=0.8cm of b](c);
     \vertex [right=0.25cm of b](c14);
     \vertex [right=0.5cm of b](c23);
     \vertex [above=0.46cm of c23](v23)[dot]{};
      \vertex [above=0.2cm of c14](v14)[]{};
    \vertex [above=0.9cm of c](j1);
    \vertex [right=0.8cm of j1](j4);
    \vertex [right=0.3cm of j4](j2);
    \vertex [right=0.5cm of j2](j3);
   	 \diagram*{(a) -- [blue, ultra thick] (b),(b) -- [thick] (j1),(v23) -- [thick] (j2),(b) -- [thick] (j4),(b)--[thick](j3)};
    \end{feynman}  
  \end{tikzpicture}={\langle \widehat\npre(\sc 1)\star \widehat\npre(\sc 2) \star \widehat\npre([\sc 3,4])\rangle \over p_{34}^2-m^2} + {\langle \widehat\npre(\sc 2)\star \widehat\npre(\sc 3) \star \widehat\npre([4,\sc 1])\rangle \over p_{14}^2-m^2} \nn\\
  &={-\langle \ck 1\star \ck 2\star \ck 4 \star\ck 3 \rangle \over p_{34}^2-m^2}+{\langle \ck 2\star \ck 3\star \ck 4 \star\ck 1 \rangle \over (p_{14}^2-m^2)}=-{p_{12}\Cdot F_4\Cdot p_3  \tr(t^{a_1}t^{a_2}t^{a_3})\over (p_{123}^2-q^2)(p_{34}^2-m^2)}+{ p_{23}\Cdot F_4\Cdot p_1  \tr(t^{a_1}t^{a_2}t^{a_3})\over (p_{123}^2-q^2) (p_{14}^2-m^2)}\, .
  \end{align}
  
 It is easy to verify that $1/(p^2_{123}-q^2)$ is a spurious pole. And  the physical poles are $1/(p_{34}^2-m^2)$ and $1/(p_{14}^2-m^2)$, on which expected factorisation behaviour can be checked. 
 The factorisation behaviour on $1/(p_{14}^2-m^2)$ is similar.  Another similar example is the five-point form factor of ${\rm Tr}(\phi^4)$
 \begin{align}
    {\cal F}(\sc{1},\sc{2}, \sc{3},\sc{4},5)&={\langle \widehat\npre(\sc 1)\star \widehat\npre(\sc 2) \star \widehat\npre(\sc 3) \star\widehat\npre([\sc 4,5])\rangle \over p_{45}^2-m^2}+{\langle \widehat\npre(\sc 2)\star \widehat\npre(\sc 3) \star \widehat\npre(\sc 4) \star\widehat\npre([\sc 5,1])\rangle \over p_{15}^2-m^2}\nn\\
 	 &=-{ p_{123}\Cdot F_5\Cdot p_4  \over (p_{1234}^2-q^2)(p_{45}^2-m^2)}+{ p_{234}\Cdot F_5\Cdot p_1  \over (p_{1234}^2-q^2) (p_{14}^2-m^2)}\, .
 \end{align}
Once again it is straightforward to verify the cancellation of the spurious poles as well as the factorisation behaviours on the physical poles. 
 
In the very end, we give examples specialised for the novel relation given in eq.~\eqref{eq:ff-relation2}. 
When $j$ line is a gluon, the identity is manifest according to the vanishing condition, e.g. 
\begin{align}
	&\langle\widehat\npre([\sc 1,4])\star \widehat\npre(\sc 2) \star\widehat\npre(\sc 3)+\widehat\npre(\sc 1)\star \widehat\npre([\sc 2,4]) \star\widehat\npre(\sc 3)+\widehat\npre(\sc 1)\star \widehat\npre(\sc 2) \star\widehat\npre([\sc 3,4])\rangle\nn\\
	&=\langle[\ck 1, \ck 4]\star \ck 2\star \ck 3+\ck 1, \star [\ck 2, \ck 4]\star \ck 3+\ck 1\star \ck 2\star [\ck 3, \ck 4]\rangle\nn\\
	&=\langle -\ck 4\star \ck 1\star \ck 2\star \ck 3+\ck 1\star \ck 2\star \ck 3\star \ck 4\rangle=0. 
\end{align} 
The identity is more non-trivial when $j$ line is a scalar, for which we have 
 \begin{align}
 	&\langle\widehat\npre([\sc 1,\sc 5])\star \widehat\npre(\sc 2) \star\widehat\npre([3,\sc 4])+\widehat\npre(\sc 1)\star \widehat\npre([\sc 2,\sc 5]) \star\widehat\npre([3,\sc 4])\rangle +\widehat\npre(\sc 1)\star \widehat\npre(\sc 2) \star\widehat\npre([[3,\sc 4],\sc 5]]) \nn\\
 	&=-\langle  \ck 5\star \ck 1\star \ck 2\star [\ck 3, \ck 4] \rangle +\langle \ck 1\star \ck 2\star [\ck 3, \ck 4] \star \ck 5\rangle =-{p_{125}\Cdot F_3\Cdot p_4\over p^2_{1245}-q^2}+{p_{12}\Cdot F_3\Cdot p_{45}\over p^2_{1245}-q^2}-{p_{124}\Cdot F_3\Cdot p_{5}\over p^2_{1245}-q^2}=0 \, .
 \end{align}


\twocolumngrid


%

\end{document}